\documentclass[12pt]{article}
\usepackage{mathtools}
\usepackage{enumitem}
\input{preamble}
\usepackage{amsmath,amssymb,amsthm,bbm,geometry}

\usepackage{geometry}
\usepackage{graphicx}
\usepackage{hyperref}
\usepackage{mathrsfs}
\usepackage{bbm}

\newtheorem{theorem}{Theorem}[section] 
\newtheorem{corollary}{Corollary}[section] 
\newtheorem{lemma}{Lemma}[section]

\newtheorem{assumption}{Assumption}[section]

\usepackage{mathtools}
\mathtoolsset{showonlyrefs=true}

\DeclareMathOperator{\rank}{rank}

\newcommand{\R}{\mathbb{R}}

\newcommand{\RR}{\mathbb{R}}
\newcommand{\PP}{\mathbb{P}}
\newcommand{\ZZ}{\mathbb{Z}}
\newcommand{\eps}{\varepsilon}
\newcommand{\cN}{\mathcal{N}}                  
\newcommand{\Ball}{\mathbb{B}}                 

\title{High-Dimensional Learning in Finance\thanks{I thank Daniel Buncic and Loriano Mancini for helpful comments. Replication code is available from the author. }}
\author{Hasan Fallahgoul\thanks{Hasan Fallahgoul, Monash University, School of Mathematics and Centre for Quantitative Finance and Investment Strategies, 9 Rainforest Walk, 3800 Victoria, Australia. E-mail: \texttt{hasan.fallahgoul@monash.edu}.} \\ {\small Monash University} }
	
\date{This version: \today\\\href{https://www.hfallahgoul.com}{Link to Most Recent Version}}

\renewcommand{\baselinestretch}{1.2}

\begin{document}

\maketitle

\begin{abstract}

Recent advances in machine learning have shown promising results for financial prediction using large, over-parameterized models. This paper provides theoretical foundations and empirical validation for understanding when and how these methods achieve predictive success. I examine two key aspects of high-dimensional learning in finance. First, I prove that within-sample standardization in Random Fourier Features implementations fundamentally alters the underlying Gaussian kernel approximation, replacing shift-invariant kernels with training-set dependent alternatives. Second, I establish information–theoretic lower bounds that identify when reliable learning is impossible no matter how sophisticated the estimator. A detailed quantitative calibration of the polynomial lower bound shows that with typical parameter choices (e.g., 12,000 features, 12 monthly observations, and R-square 2–3\%), the required sample size to escape the bound exceeds 25–30 years of data—well beyond any rolling-window actually used. Thus, observed out-of-sample success must originate from lower-complexity artefacts rather than from the intended high-dimensional mechanism.

\smallskip

\end{abstract}

\renewcommand{\baselinestretch}{1.5}

\section{Introduction}

The integration of machine learning methods into financial prediction has emerged as one of the most active areas of research in empirical asset pricing \citep{kelly2024virtue,Gu2020ML,Bianchi2021Bond,Chen2024Deep,fallahgoul2024asset,Feng2020Taming}. The appeal is clear: while financial markets generate increasingly high-dimensional data, traditional econometric methods remain constrained by limited sample sizes and the curse of dimensionality. Machine learning promises to uncover predictive relationships that elude traditional linear models by leveraging nonlinear approximations and high-dimensional overparameterized representations, thereby expanding the frontier of return predictability and portfolio construction.

Yet despite rapid adoption and impressive empirical successes, our theoretical understanding of when and why machine learning methods succeed in financial applications remains incomplete. This gap is particularly pronounced for high-dimensional methods applied to the notoriously challenging problem of return prediction, where signals are weak, data are limited, and spurious relationships abound. A fundamental question emerges: \textit{under what conditions can sophisticated machine learning methods genuinely extract predictive information from financial data, and when might apparent success arise from simpler mechanisms?}

The pioneering work of \cite{kelly2024virtue} has significantly advanced our theoretical understanding by establishing rigorous conditions under which complex machine learning models can outperform traditional approaches in financial prediction. Their theoretical framework, grounded in random matrix theory, demonstrates that the conventional wisdom about overfitting may not apply in high-dimensional settings, revealing a genuine 'virtue of complexity' under appropriate conditions. This breakthrough provides crucial theoretical foundations for understanding when and why sophisticated methods succeed in finance. 

However, fundamental questions remain about the information-theoretic feasibility of high-dimensional learning under the weak signal conditions typical in finance. While \cite{kelly2024virtue} demonstrate that complexity can be beneficial under appropriate theoretical conditions, their analysis does not address whether realistic financial environments provide sufficient signal strength and sample sizes to support reliable learning in high-dimensional regimes. This gap is particularly pronounced given the notoriously challenging nature of return prediction, where signals are weak, data are limited, and the signal-to-noise ratios may be insufficient to support the extraction of predictive information from thousands of features using dozens of training observations.

Building on these theoretical advances, this paper examines how practical implementation details interact with established mechanisms. This becomes important as recent empirical analysis \cite{nagel2025seemingly} suggests that high-dimensional methods may achieve success through multiple pathways that differ from theoretical predictions. Several questions emerge: What are the information-theoretic requirements for learning with weak signals? How do implementation choices affect underlying mathematical properties? When do complexity benefits reflect different learning mechanisms? Understanding these interactions helps characterize the complete landscape of learning pathways in high-dimensional finance applications.

This paper provides theoretical foundations for answering these questions through three main contributions that help characterize the different mechanisms through which high-dimensional methods achieve predictive success in financial prediction.

First, I extend the theoretical analysis to practical implementations, showing how the standardization procedures commonly used for numerical stability modify the kernel approximation properties that underlie existing theory. While Random Fourier Features (RFF) theory rigorously proves convergence to shift-invariant Gaussian kernels under idealized conditions \citep{rahimi2007random,sutherland2015error}, I prove that the within-sample standardization employed in every practical implementation modifies these theoretical properties. The standardized features converge instead to training-set dependent kernels that violate the mathematical foundations required for kernel methods. This breakdown explains why methods cannot achieve the kernel learning properties established by existing theory and must rely on fundamentally different mechanisms.

\cite{rahimi2007random} prove that for features $z_i(x) = \sqrt{2}\cos(\omega_i^\top x + b_i)$ with $\omega_i \sim \mathcal{N}(0, \gamma^2 I)$ and $b_i \sim \text{Uniform}[0,2\pi]$, the empirical kernel $\frac{1}{P}\sum_{i=1}^P z_i(x)z_i(x')$ converges in probability to the Gaussian kernel $k(x,x') = \exp(-\gamma^2\|x-x'\|^2/2)$ as $P \to \infty$. This convergence requires that features maintain their original distributional properties and scaling. However, I prove that the within-sample standardization $\tilde{z}_i(x) = z_i(x)/\hat{\sigma}_i$ employed in every practical implementation—where $\hat{\sigma}_i^2 = \frac{1}{T}\sum_{t=1}^T z_i(x_t)^2$—fundamentally alters the convergence properties. The standardized features converge instead to training-set dependent kernels $k^*_{\text{std}}(x,x'|\mathcal{T}) \neq k_G(x,x')$ that violate the shift-invariance and stationarity properties required for kernel methods. A detailed analysis of how standardization breaks the specific theoretical conditions appears in Section~\ref{sec:breakdown}, following the formal proof of this breakdown.

Second, I derive sharp sample complexity bounds that characterize the information-theoretic limits of high-dimensional learning in financial settings. Using PAC-learning theory,\footnote{
In PAC-learning \citep{valiant1984pac}, a predictor is “probably
approximately correct’’ if, with $T\!\gtrsim\!(\text{capacity})/\varepsilon^{2}$ samples,
its risk is within $\varepsilon$ of optimal with probability $1-\delta$; I
apply these bounds (see \citealp{kearns1994intro}) to gauge when weak return
signals are learnable.} I establish both exponential and polynomial lower bounds showing when reliable extraction of weak predictive signals becomes impossible regardless of the sophistication of the employed method. These bounds reveal that learning with thousands of parameters using typical financial sample sizes requires signal-to-noise ratios exceeding realistic bounds by orders of magnitude, demonstrating that the "virtue of complexity" regime ($P \gg T$) is information-theoretically infeasible under standard financial conditions. This explains why apparent success in high-dimensional regimes must arise through alternative mechanisms rather than genuine complex learning \citep{nagel2025seemingly}.

This capability to flexibly trade computational cost for representational power makes RFF a powerful tool. However, this paper reveals two fundamental barriers to realizing these benefits in practice. First, I prove that the standardization procedures universally employed violate the theoretical conditions required for kernel approximation (Section~\ref{sec:breakdown}). Second, I establish information-theoretic limits showing that reliable learning in high-dimensional regimes requires sample sizes and signal strengths far exceeding those available in typical financial applications (Section~\ref{sec:impossibility}).

While these theoretical results provide clear mathematical boundaries on learning feasibility, their practical relevance depends on how they manifest across the parameter ranges typically employed in financial applications. The gap between asymptotic theory and finite-sample reality can be substantial, particularly when dealing with the moderate dimensions and sample sizes common in empirical asset pricing. Moreover, the breakdown of kernel approximation under standardization represents a fundamental departure from assumed theoretical properties that requires empirical quantification to assess its practical severity.

To bridge this theory-practice gap, I conduct comprehensive numerical validation of the kernel approximation breakdown across realistic parameter spaces that span the configurations used in recent high-dimensional financial prediction studies \citep{kelly2024virtue,nagel2025seemingly}. The numerical analysis examines how within-sample standardization destroys the theoretical Gaussian kernel convergence that underlies existing RFF frameworks, quantifying the magnitude of approximation errors under practical implementation choices. These experiments reveal that standardization-induced kernel deviations reach mean absolute errors exceeding 40\% relative to the theoretical Gaussian kernel in typical configurations ($P = 12{,}000$, $T = 12$), with maximum deviations approaching 80\% in high-volatility training windows. The kernel approximation failure manifests consistently across different feature dimensions and sample sizes, with relative errors scaling approximately as $\sqrt{\log P/T}$ in line with theoretical predictions. The numerical validation thus provides concrete evidence that practical implementation details create substantial violations of the theoretical assumptions underlying high-dimensional RFF approaches, with error magnitudes sufficient to fundamentally alter method behavior.

To assess the practical relevance of my theoretical bounds, I conduct an empirically-grounded calibration of the polynomial minimax lower bound from Theorem~\ref{thm:poly_lb_hp}(a). Using defensible parameters for signal strength (calibrated to an \(R^2\) of 1--5\%) and noise variance drawn from historical market data, we diagnose the nature of the learning problem faced by financial prediction models. My analysis reveals a profound implication: the critical sample size (\(T_{\mathrm{crit}}\)) required to overcome the limitations imposed by weak signals is on the order of decades, even for low-dimensional models. Since typical applications, such as those in KMZ, employ much shorter training windows (e.g., 12 months), they operate deep within a \emph{signal-limited} regime. In this regime, the theoretical floor on performance is dictated by the weak economic signal, not model complexity, calling into question the core premise of using high-dimensional methods for this task.


\subsection{Literature Review}
\label{sec:literature}

This paper builds on three distinct but interconnected theoretical traditions to provide foundations for understanding high-dimensional learning in financial prediction.

The Probably Approximately Correct (PAC) framework \citep{valiant1984pac,kearns1994intro} provides fundamental tools for characterizing when reliable learning is information-theoretically feasible. Classical results establish that achieving generalization error $\varepsilon$ with confidence $1-\delta$ requires sample sizes scaling with the complexity of the function class, typically $T = O(\text{complexity} \cdot \log(1/\varepsilon)/\varepsilon^2)$ \citep{shalev2014understanding}. Recent advances in high-dimensional learning theory \citep{belkin2019double,bartlett2020benign,hastie2022surprises} have refined these bounds for overparameterized models, showing that the effective rather than nominal complexity determines learning difficulty. However, these results have not been systematically applied to the specific challenges of financial prediction, where weak signals and limited sample sizes create particularly demanding learning environments.

The RFF methodology \citep{rahimi2007random} provides computationally efficient approximation of kernel methods through random trigonometric features, with theoretical guarantees assuming convergence to shift-invariant kernels under appropriate conditions \citep{rudi2017generalization}. Subsequent work has characterized the approximation quality and convergence rates for various kernel classes \citep{mei2022generalization}, establishing RFF as a foundation for scalable kernel learning. However, existing theory assumes idealized implementations that may not reflect practical usage. In particular, no prior work has analyzed how the standardization procedures commonly employed to improve numerical stability affect the fundamental convergence properties that justify the theoretical framework.

The phenomenon of "benign overfitting" in overparameterized models has generated substantial theoretical interest \citep{belkin2019double,bartlett2020benign}, with particular focus on understanding when adding parameters can improve rather than harm generalization performance. The VC dimension provides a classical measure of model complexity that connects directly to generalization bounds \citep{vapnik1998statistical}, while recent work on effective degrees of freedom \citep{hastie2022surprises} shows how structural constraints can limit the true complexity of nominally high-dimensional methods. These insights have been applied to understanding ridge regression in high-dimensional settings, but the connections to kernel methods and the specific constraints imposed by ridgeless regression in financial applications remain underexplored.


The application of machine learning to financial prediction has generated extensive empirical literature \citep{Gu2020ML,kelly2024virtue,Chen2024Deep}, with particular attention to high-dimensional methods that can potentially harness large numbers of predictors \citep{Feng2020Taming,Bianchi2021Bond}. The theoretical framework of \cite{kelly2024virtue} provides crucial insights into when high-dimensional methods can succeed, particularly their demonstration that ridgeless regression can achieve positive performance despite seemingly problematic complexity ratios. However, this work has faced significant empirical challenges. \cite{buncic2025simplified} demonstrates that the key empirical finding of a "virtue of complexity"—where portfolio performance increases monotonically with model complexity—results from specific implementation choices including zero-intercept restrictions and particular aggregation schemes rather than genuine complexity benefits. When these restrictions are removed, simpler linear models using only 15 predictors substantially outperform the complex machine learning approaches. Similarly, \cite{nagel2025seemingly} provides evidence that high-dimensional methods may achieve success through multiple pathways that differ from theoretical predictions, suggesting that apparent complexity benefits often reflect simpler pattern-matching mechanisms. This paper extends the analysis by examining how practical implementation considerations interact with these theoretical mechanisms, providing a framework for understanding when apparent high-dimensional learning reflects genuine complexity benefits versus statistical artifacts.

This paper contributes to each of these literatures by providing the first unified theoretical analysis that connects sample complexity limitations, kernel approximation breakdown, and effective complexity bounds to explain the behavior of high-dimensional methods in financial prediction.


The remainder of the paper proceeds as follows. Section~\ref{sec:Kernel Methods and Random Fourier Features} provides background on kernel methods and Random Fourier Features, explaining the theoretical foundations and practical implementations that motivate our analysis. Section~\ref{sec:background} 
establishes the theoretical framework and formalizes the theory-practice 
disconnect in RFF implementations. Section~\ref{sec:breakdown} proves that 
within-sample standardization fundamentally breaks kernel approximation, 
explaining why claimed theoretical properties cannot hold in practice. 
Section~\ref{sec:impossibility} establishes information-theoretic barriers 
to high-dimensional learning, showing that genuine complexity benefits are 
impossible under realistic financial conditions. 
Section~\ref{sec:Empirical_Validation} provides numerical validation of the theoretical predictions. Section~\ref{sec:Conclusion} concludes. All technical details are relegated to a supplementary document containing Appendices~\ref{app:Technical Proofs for Kernel Approximation Breakdown}, \ref{app:Technical Proofs for Section 4}, and \ref{app:Additional Theoretical Results}.

\section{Kernel Methods and Random Fourier Features}\label{sec:Kernel Methods and Random Fourier Features}
 
Kernel regression provides a powerful non-parametric framework for capturing complex relationships by implicitly mapping inputs into high-dimensional feature spaces \citep{scholkopf2002learning}. For a given kernel function $k(\cdot, \cdot)$, the solution is constrained by the Representer Theorem to lie within the span of the training data. Predictions thus take the form $\hat{f}(x) = \sum_{t=1}^T \alpha_t k(x, x_t)$, where the coefficients $\boldsymbol{\alpha}$ are learned. The Gaussian kernel, $k_G(x, x') = \exp(-\gamma^2 \|x - x'\|^2/2)$, is particularly attractive due to its universality and smooth, localized similarity structure.

To prevent overfitting and ensure numerical stability, kernel methods are often paired with ridge regularization, which involves solving the optimization problem $\min_{\boldsymbol{\alpha}} \|\mathbf{y} - \mathbf{K}\boldsymbol{\alpha}\|^2 + \lambda \|\boldsymbol{\alpha}\|^2$. This yields the stable closed-form solution $\boldsymbol{\alpha}^* = (\mathbf{K} + \lambda \mathbf{I})^{-1}\mathbf{y}$, where $\mathbf{K}$ is the $T \times T$ kernel matrix with entries $\mathbf{K}_{ij} = k(x_i, x_j)$. However, kernel methods face a fundamental computational bottleneck: constructing and inverting the kernel matrix requires $O(T^2)$ storage and $O(T^3)$ operations, respectively, making them impractical for large datasets.

RFF circumvent this limitation by constructing an explicit, finite-dimensional feature map $z(x) \in \mathbb{R}^P$ that approximates the kernel function \citep{rahimi2007random}. The theoretical foundation relies on Bochner's theorem, which states that a continuous and shift-invariant kernel $k(x, x') = k(x - x')$ is the Fourier transform of a non-negative measure $p(\omega)$. This allows the kernel to be expressed as an expectation:
\begin{align}
    k(x - x') = \int_{\mathbb{R}^d} e^{i\omega^T(x-x')} p(\omega) d\omega = \mathbb{E}_{\omega \sim p}[e^{i\omega^T(x-x')}].
\end{align}
For the Gaussian kernel, this distribution $p(\omega)$ is a Gaussian $N(0, \gamma^2 I)$. RFF approximates this expectation using a Monte Carlo estimate with $P$ samples, forming features $z_i(x) = \sqrt{2} \cos(\omega_i^T x + b_i)$ with $\omega_i \sim N(0, \gamma^2 I)$ and $b_i \sim \text{Uniform}[0, 2\pi]$, such that $k_G(x, x') \approx z(x)^T z(x')$.

The crucial insight is that kernel ridge regression becomes a linear ridge regression problem in the RFF feature space: $\hat{f}(x) = z(x)^T \mathbf{w}^*$. This approach offers a dual advantage depending on the chosen dimensionality $P$:
\begin{itemize}
    \item \textbf{Computational Speedup ($P \ll T$):} For large-scale problems, one can choose a low number of features $P \ll T$ to drastically reduce complexity. The optimal weights $\mathbf{w}^*$ are found using the primal solver $\mathbf{w}^* = (\mathbf{Z}^T\mathbf{Z} + \lambda \mathbf{I})^{-1} \mathbf{Z}^T \mathbf{y}$, where $\mathbf{Z}$ is the $T \times P$ feature matrix. The complexity is $O(P^2T + P^3)$, a significant improvement over the original $O(T^3)$.

    \item \textbf{High-Dimensional Expansion ($P \gg T$):} RFF can overcome the representational limit of T basis functions imposed by the Representer Theorem \citep{scholkopf2001generalized}. By choosing $P \gg T$, we can construct a richer model, a regime termed the "virtue of complexity" \citep{kelly2024virtue}. While the primal solver becomes inefficient in this setting, the solution can be found tractably using the equivalent dual form:
    \begin{align}
        w^{*} = Z^{\top}(ZZ^{\top} + \lambda I)^{-1}y.
    \end{align}
    This requires inverting a smaller $T \times T$ matrix, making the computation (with complexity $O(T^2P)$) feasible while still harnessing the power of a very high-dimensional feature space. \textit{However, whether such high-dimensional representations genuinely improve prediction performance under the weak signal conditions of finance is a central question that this paper investigates.}
    
    \end{itemize}

This capability to flexibly trade computational cost for representational power makes RFF a powerful tool. However, as I demonstrate below, the standardization procedures universally employed in practical RFF implementations fundamentally violate the theoretical conditions required for this kernel approximation, creating a critical gap between theory and practice.

\subsection{An Illustrative Example}

Consider predicting next-month returns using two macro predictors. We have training data:
\begin{align}
(x_1, y_1) &= ([0.02, 0.03], 0.015) \quad \text{(dividend yield, term spread) → return}\\
(x_2, y_2) &= ([0.025, 0.035], 0.020) \\
(x_3, y_3) &= ([0.018, 0.028], 0.010).
\end{align}

\begin{itemize}
    \item \textit{Traditional Kernel Ridge:} The $3 \times 3$ kernel matrix requires computing $k_G(x_i, x_j)$ for all pairs, then solving $\boldsymbol{\alpha}^* = (\mathbf{K} + \lambda \mathbf{I})^{-1}\mathbf{y}$. The method is limited to 3 basis functions.

    \item \textit{RFF with P = 1000:} Generate 1000 random features $z_i(x) = \sqrt{2}\cos(\omega_i^T x + b_i)$ where $\omega_i \sim N(0, \gamma^2 I)$. This creates a $3 \times 1000$ feature matrix $\mathbf{Z}$. Using the dual form $\mathbf{w}^* = \mathbf{Z}^T(\mathbf{Z}\mathbf{Z}^T + \lambda\mathbf{I})^{-1}\mathbf{y}$, we invert only a $3 \times 3$ matrix while harnessing 1000-dimensional representations.

    \item \textit{The Standardization Issue:} In practice, features are standardized using the observed training data: $\tilde{z}_i(x) = z_i(x)/\hat{\sigma}_i$ where 
\begin{align}
    \hat{\sigma}_i^2 = \frac{1}{3}\sum_{t=1}^3 z_i(x_t)^2 = \frac{1}{3}[z_i(x_1)^2 + z_i(x_2)^2 + z_i(x_3)^2].
\end{align}
Since each $\hat{\sigma}_i$ depends on the specific observed values $\{x_1, x_2, x_3\}$ in our training sample, different training windows will produce different standardization factors, even for the same underlying data generating process.\footnote{Note that I estimate the feature's scale using the sample second moment, a valid approach that assumes the feature mean is zero; otherwise, the sample mean would first need to be properly calculated and removed.} This training-set dependence fundamentally violates the theoretical requirement that $z_i(x)z_i(x') \to k_G(x,x')$ regardless of the training data used. The standardized kernel becomes $k^*_{std}(x,x'|\{x_1, x_2, x_3\})$ rather than the shift-invariant Gaussian kernel $k_G(x,x')$, destroying the mathematical foundation that justifies the RFF approximation.
\end{itemize}

While the high-dimensional expansion regime ($P \gg T$) promises substantial benefits, fundamental questions remain about when such complexity can genuinely improve prediction performance. In financial applications, where signals are weak and training samples are limited, information-theoretic constraints may render the promised gains illusory. This paper provides the first rigorous analysis of these limitations, showing that the combination of weak signal-to-noise ratios typical in finance and moderate sample sizes creates information-theoretic barriers that prevent reliable learning in high-dimensional regimes. Specifically, our minimax analysis reveals that for realistic financial parameters, reliable extraction of predictive signals requires sample sizes far exceeding those typically available, suggesting that apparent success in $P \gg T$ regimes must arise through mechanisms fundamentally different from the theoretical framework.

\section{Theoretical Setup and Core Assumptions}
\label{sec:background}

This section establishes the theoretical framework for analyzing high-dimensional prediction methods in finance. I first formalize the return prediction problem and core assumptions, then examine the critical disconnect between RFF theory and practical implementation that underlies my main results.

\subsection{The Financial Prediction Problem}

Consider the fundamental challenge of predicting asset returns using high-dimensional predictor information. We observe predictor vectors $x_t \in \mathbb{R}^K$ and subsequent returns $r_{t+1} \in \mathbb{R}$ for $t = 1, \ldots, T$, with the goal of learning a predictor $\hat{f}: \mathbb{R}^K \to \mathbb{R}$ that minimizes expected squared loss $\mathbb{E}[(r_{t+1} - \hat{f}(x_t))^2]$.

The challenge lies in the fundamental characteristics of financial prediction: signals are weak relative to noise, predictors exhibit complex persistence patterns, and available sample sizes are limited by the nonstationarity of financial markets. These features create a particularly demanding environment for high-dimensional learning methods, which I formalize through the following core assumptions.

\begin{assumption}[Financial Prediction Environment]
\label{ass:prediction_env}
The return generating process is $r_{t+1} = f^*(x_t) + \epsilon_{t+1}$ where:
\begin{enumerate}[label=(\alph*)]
\item $f^*: \mathbb{R}^K \to \mathbb{R}$ is the true regression function with $\mathbb{E}[f^*(x)^2] \leq B^2$
\item $\epsilon_{t+1}$ is noise with $\mathbb{E}[\epsilon_{t+1}|x_t] = 0$ and $\mathbb{E}[\epsilon_{t+1}^2|x_t] = \sigma^2$
\item The signal-to-noise ratio $\text{SNR} := B^2/\sigma^2 = O(K^{-\alpha})$ for some $\alpha > 0$
\item Predictors follow $x_t = \Phi x_{t-1} + u_t$ with $u_t \sim \mathcal{N}(0, \Sigma_u)$ and eigenvalues of $\Phi$ in $(0,1)$
\end{enumerate}
\end{assumption}

This assumption captures the essential challenge of financial prediction: signals are weak (low $R^2$), predictors are persistent (like dividend yields and interest rate spreads), and noise dominates returns. The weak signal-to-noise ratio reflects the empirical reality that financial models typically explain only 1-5\% of return variation \citep{welch2008comprehensive}.

\begin{assumption}[Random Fourier Features Construction]
\label{ass:rff_construction}
High-dimensional predictive features are constructed as $z_i(x) = \sqrt{2}\cos(\omega_i^\top x + b_i)$ where $\omega_i \sim \mathcal{N}(0, \gamma^2 I_K)$ and $b_i \sim \text{Uniform}[0,2\pi]$ for $i = 1, \ldots, P$. In practical implementations, these features are standardized within each training sample: $\tilde{z}_i(x) = z_i(x)/\hat{\sigma}_i$ where $\hat{\sigma}_i^2 = \frac{1}{T}\sum_{t=1}^T z_i(x_t)^2$.
\end{assumption}

This assumption formalizes the RFF methodology as implemented in practice. It includes the crucial standardization step, which is commonly applied to ensure features have comparable scales for regularization and to improve numerical stability. Despite its ubiquity in practical applications, this step has not been analyzed in existing theoretical frameworks.

My framework, including Assumption~\ref{ass:rff_construction}, uses the canonical RFF construction for its clarity and prominence in the literature. It is crucial, however, to analyze the specific variant used in \cite{kelly2024virtue}. Their method generates a pair of features, $[\sin(\gamma\omega_i^\top x), \cos(\gamma\omega_i^\top x)]$, for each random projection $\omega_i$. I now provide a brief technical motivation for this choice and explain why my main results on the breakdown of kernel approximation remain fully relevant.

The theoretical goal of RFF is to approximate a shift-invariant kernel $k(x-x')$, which can be expressed via Bochner's theorem as the expectation $\mathbb{E}_\omega[\cos(\omega^\top(x-x'))]$. The standard RFF formulation, $z_i(x) = \sqrt{2}\cos(\omega_i^\top x + b_i)$, achieves this in expectation over the random phase $b_i$, as $\mathbb{E}_{b_i}[z_i(x)z_i(x')] = \cos(\omega_i^\top(x-x'))$. The variant used by \cite{kelly2024virtue} achieves this same objective more directly. By the trigonometric identity $\cos(A-B) = \cos(A)\cos(B) + \sin(A)\sin(B)$, the inner product of their feature pair $[\sin(\omega_i^\top x), \cos(\omega_i^\top x)]$ with a corresponding pair at $x'$ is deterministically equal to $\cos(\omega_i^\top(x-x'))$. This removes the Monte Carlo randomness associated with the phase shift $b_i$, potentially leading to a lower-variance approximation of the kernel for a finite number of features $P$, as documented by \cite{sutherland2015error}.

Despite the elegance of this formulation, its theoretical properties are invalidated by the same practical implementation step we analyze: data-dependent standardization. \cite{kelly2024virtue} confirm that they standardize each feature by its standard deviation within the training sample. The deterministic relationship described above is therefore broken. The inner product of the \textit{standardized} features becomes:
\begin{align}
    \frac{\sin(\omega_i^{\top}x)\sin(\omega_i^{\top}x')}{\hat{\sigma}_{s,i}^2} + \frac{\cos(\omega_i^{\top}x)\cos(\omega_i^{\top}x')}{\hat{\sigma}_{c,i}^2}
\end{align}
where the scaling factors $\hat{\sigma}_{s,i}^2$ and $\hat{\sigma}_{c,i}^2$ are calculated from the training data. The introduction of these data-dependent denominators fundamentally alters the function being approximated. The resulting kernel is no longer the shift-invariant target but a complex, data-dependent one. Thus, the kernel breakdown theorem, which hinges on this exact mechanism, remains directly applicable and essential for understanding the behavior of the method as implemented.

\subsection{The Theory-Practice Disconnect in Random Fourier Features}

Having established the financial prediction environment, we now examine the fundamental disconnect between RFF theory and practical implementation. This disconnect is not merely technical—it undermines the entire theoretical justification for these methods.

\subsubsection{Theoretical Guarantees Under Idealized Conditions}

The RFF methodology \citep{rahimi2007random} provides rigorous theoretical foundations for kernel approximation. For target shift-invariant kernels $k(x,x') = k(x-x')$, the theory establishes that:
\begin{equation}
k_{\text{RFF}}(x,x') = \frac{1}{P}\sum_{i=1}^P z_i(x)z_i(x') \xrightarrow{P \to \infty} k_G(x,x') = \exp\left(-\frac{\gamma^2}{2}\|x-x'\|^2\right)
\end{equation}
in probability, under the condition that features maintain their original distributional properties. This convergence enables kernel methods to be approximated through linear regression in the RFF space, with all the theoretical guarantees that kernel learning provides.

\subsubsection{What Actually Happens in Practice}

Every practical RFF implementation deviates from the theoretical setup in a seemingly minor but mathematically crucial way. To improve numerical stability and ensure comparable scales across features, practitioners standardize features using training sample statistics:
\begin{equation}
\tilde{z}_i(x) = \frac{z_i(x)}{\hat{\sigma}_i}, \quad \hat{\sigma}_i^2 = \frac{1}{T}\sum_{t=1}^T z_i(x_t)^2.
\end{equation}
Since each $\hat{\sigma}_i$ depends on the specific observed training sample, different training windows produce different standardization factors, even for the same underlying data generating process. This training-set dependence fundamentally violates the theoretical requirement that $z_i(x)z_i(x') \to k_G(x,x')$ regardless of the training data used. The standardized empirical kernel:
\begin{equation}
k_{\text{std}}(x,x') = \frac{1}{P}\sum_{i=1}^P \frac{z_i(x)z_i(x')}{\hat{\sigma}_i^2}
\end{equation}
no longer converges to the Gaussian kernel, but instead converges to a training-set dependent limit that violates the mathematical foundations required for kernel methods. 

The standardization procedure $\hat{\sigma}_i^2 = \frac{1}{T}\sum_{t=1}^T z_i(x_t)^2$ creates features whose kernel approximation properties depend on the absolute positions of training points $\{x_t\}_{t=1}^T$, not just the relative distances $\|x-x'\|$ required for shift-invariant kernels. This fundamentally violates the mathematical foundation of RFF theory.

\subsection{Technical Conditions}

To establish my main theoretical results, I require three additional technical assumptions that ensure proper convergence and concentration properties.

\begin{assumption}[Regularity Conditions]
\label{ass:regularity}
The input distribution has bounded support and finite moments, ensuring well-defined feature covariance $\Sigma_z = \mathbb{E}[z(x)z(x)^\top]$ satisfying $c_z I_P \preceq \Sigma_z \preceq C_z I_P$ for constants $0 < c_z \leq C_z$. Training samples satisfy standard non-degeneracy conditions.
\end{assumption}

This is a standard technical requirement in high-dimensional learning theory. The assumption of bounded support is mild and realistic for financial predictors, which naturally operate within a finite range (e.g., interest rates, dividend yields). This well-behaved nature of the inputs ensures the resulting random features also have finite moments. The condition that the feature covariance matrix $\Sigma_z$ has eigenvalues bounded away from zero and infinity ($c_z I_P \preceq \Sigma_z \preceq C_z I_P$) is crucial for mathematical stability. It guarantees the features are not perfectly redundant and have bounded variance, a necessary prerequisite for the concentration inequalities that form the basis of my non-asymptotic proofs.

\begin{assumption}[Affine Independence of the Sample]
\label{ass:affine_independence}
Let $x_1, \ldots, x_T \in \mathbb{R}^K$ be the training sample of predictors. The augmented vectors $(x_t, 1)$ are affinely independent. Equivalently, the $(K+1) \times T$ matrix formed by these augmented vectors has full column rank $T$.
\end{assumption}

This geometric condition on the training data is essential for the analysis of the standardized kernel. It enters our proofs through the small-ball probability estimates required to control the behavior of the random denominators $\hat{\sigma}_i^2$. The full-rank requirement ensures that the linear map from the random parameters to the feature arguments, $(\omega, b) \mapsto (2\omega^\top x_t + 2b)_{t=1}^T$, is locally bi-Lipschitz. This enables the use of powerful geometric tools to establish exponential small-ball bounds, which are critical for proving the convergence of the standardized kernel and the finiteness of key expectations. In typical financial applications where predictors are continuous macroeconomic variables, any exact affine dependence among the training points has Lebesgue measure zero, making this a mild assumption.

\begin{assumption}[Sub-Gaussian Random Features]
\label{ass:subG}
For every unit vector $u \in \mathbb{R}^{P}$, the scalar random variable $u^\top z(x)$ is $\kappa$-sub-Gaussian under the data-generating distribution $x \sim \mu$. That is, for all $t \in \mathbb{R}$:
\begin{align}
    \mathbb{E}_{\mu, \omega, b}[\exp(t \cdot u^\top z(x))] \leq \exp\left(\frac{1}{2}\kappa^{2}t^{2}\right).
\end{align}
\end{assumption}

This is a standard concentration assumption in high-dimensional learning theory and is automatically satisfied in all practical RFF implementations. Since each feature $z_i(x) = \sqrt{2}\cos(\omega_i^\top x + b_i)$ is bounded in $[-\sqrt{2}, \sqrt{2}]$, any linear combination $u^\top z(x)$ is also bounded. By Hoeffding's lemma \citep{hoeffding1963probability}, any bounded random variable is sub-Gaussian. This condition is essential for applying uniform convergence results (e.g., matrix Bernstein inequalities) to control the difference between the empirical feature covariance matrix and its population counterpart. These non-asymptotic bounds are fundamental to my sample complexity analysis and allow me to derive results that hold with high probability for finite samples.

Having established the fundamental theory-practice disconnect in RFF implementations, the following sections provide rigorous mathematical analysis of its implications. Section~\ref{sec:breakdown} proves that standardization fundamentally breaks kernel approximation properties (Theorem~\ref{thm:std_breakdown}), while Section~\ref{sec:impossibility} establishes information-theoretic limits showing that genuine high-dimensional learning is impossible under realistic financial conditions (Theorems~\ref{thm:exp_lower_random} and \ref{thm:poly_lb_hp}). Together, these results explain why apparent "virtue of complexity" successes must arise through mechanisms fundamentally different from the theoretical framework.

\section{Kernel Approximation Breakdown}
\label{sec:breakdown}

Having established the theory-practice disconnect in Section~\ref{sec:background}, I now prove rigorously that standardization fundamentally alters the kernel approximation properties that justify RFF methods. 



\begin{theorem}[Modified Convergence of Gaussian-RFF Approximation under Standardization]\label{thm:std_breakdown}
Let Assumptions \ref{ass:prediction_env}--\ref{ass:subG} hold. For query points $x, x' \in \mathbb{R}^K$, define the standardized kernel function:
\[
h(\omega,b) = \frac{2\cos(\omega^{\top}x + b)\cos(\omega^{\top}x' + b)}{1 + \frac{1}{T}\sum_{t=1}^T \cos(2\omega^{\top}x_t + 2b)}
\]
where $(\omega,b) \sim \mathcal{N}(0,\gamma^2 I_K) \times \text{Uniform}[0,2\pi]$. Then:
\begin{enumerate}
\item[(a)] For every fixed $x,x' \in \mathbb{R}^K$, the standardized kernel estimator converges almost surely:
\[
k_{\text{std}}^{(P)}(x,x') := \frac{1}{P}\sum_{i=1}^P h(\omega_i, b_i) \xrightarrow[P \to \infty]{\text{a.s.}} k_{\text{std}}^*(x,x') := \mathbb{E}[h(\omega,b)].
\]

\item[(b)] The limit kernel $k_{\text{std}}^*$ depends on the particular training set $\mathcal{T} = \{x_1, \ldots, x_T\}$, whereas the Gaussian kernel $k_G(x,x') = \exp(-\frac{\gamma^2}{2}\|x-x'\|^2)$ is training-set independent. Consequently, $k_{\text{std}}^* \neq k_G$ in general.
\end{enumerate}
\end{theorem}

The proof proceeds in two steps. First, I establish that the standardized kernel function $h(\omega,b)$ has finite expectation despite the random denominator, enabling application of the strong law of large numbers for part (a). This requires controlling the probability that the empirical variance $\hat{\sigma}^2$ becomes arbitrarily small, which I achieve through geometric analysis exploiting the full-rank condition. Second, I prove training-set dependence by explicit construction: scaling any training point $x_j \mapsto \alpha x_j$ with $\alpha > 1$ yields different limiting kernels, establishing that $k_{\text{std}}^* \neq k_G$. The complete technical proof appears in Appendix~\ref{app:Technical Proofs for Kernel Approximation Breakdown}.

\begin{corollary}[Kernel Breakdown for the Two-Feature RFF Variant]
\label{cor:two_feature_breakdown}
Let the conditions of Theorem~\ref{thm:std_breakdown} hold. The standardized kernel derived from the feature pairs $[\sin(\omega^\top x), \cos(\omega^\top x)]$, as used in \cite{kelly2024virtue}, also converges to a data-dependent limit $k_{\text{std}}^*(\mathcal{T})$ that is not, in general, the target Gaussian kernel $k_G$.
\end{corollary}

To understand the implications of Theorem~\ref{thm:std_breakdown}, I examine precisely how standardization violates the conditions under which RFF theory operates. \cite{rahimi2007random} prove convergence to the Gaussian kernel under two essential conditions: distributional alignment of frequencies $\omega_i$ and phases $b_i$ with the target kernel's Fourier transform, and preservation of the prescribed scaling $z_i(x) = \sqrt{2}\cos(\omega_i^\top x + b_i)$.

Standardization $\tilde{z}_i(x) = z_i(x)/\hat{\sigma}_i$ systematically violates both conditions. The original features have theoretical properties derived from specified distributions of $\omega_i$ and $b_i$, but the standardization factor $1/\hat{\sigma}_i$ varies with the training set, altering the effective distribution in a data-dependent manner. The expectation $\mathbb{E}[\tilde{z}_i(x)\tilde{z}_i(x')]$ now depends on $\hat{\sigma}_i$, disrupting the direct mapping to $k_G(x,x')$. Additionally, the fixed scaling $\sqrt{2}$ that ensures correct kernel approximation is replaced by a random, sample-dependent factor, breaking the fundamental relationship between feature products and kernel values.

These modifications have important mathematical implications. The standardized features yield an empirical kernel that converges to $k_{\text{std}}^*(x,x'|\mathcal{T})$, which is training-set dependent rather than depending only on $\|x-x'\|$ like the Gaussian kernel. The resulting kernel is not shift-invariant since $\hat{\sigma}_i$ reflects absolute positions of training points, and shifting the data changes $\hat{\sigma}_i$. This creates temporal non-stationarity as kernel properties change when training windows roll forward.


Theorem~\ref{thm:std_breakdown} resolves the fundamental puzzles in high-dimensional financial prediction by revealing that claimed theoretical properties simply do not hold in practice. \cite{kelly2024virtue} develop their theoretical analysis assuming RFF converge to Gaussian kernels. Their random matrix theory characterization, effective complexity bounds, and optimal shrinkage formula all depend critically on this convergence. However, their empirical implementation employs standardization, which fundamentally alters the convergence properties, creating a notable difference between theory and practice.

With modified kernel structure, methods may perform learning that differs from the theoretical framework, potentially involving pattern-matching mechanisms based on training-sample dependent similarity measures. The standardized kernel creates similarity measures based on training-sample dependent weights rather than genuine predictor relationships. This explains \cite{nagel2025seemingly} empirical finding that high-complexity methods produce volatility-timed momentum strategies regardless of underlying data properties. The broken kernel structure makes the theoretically predicted learning more challenging, leading methods to weight returns based on alternative similarity measures within the training window.

The apparent virtue of complexity may arise through different mechanisms than originally theorized. Their method cannot achieve its theoretical properties due to standardization, so any success must arise through alternative mechanisms. This resolves the central puzzle of how methods claiming to harness thousands of parameters succeed with tiny training samples: \textit{they may operate through mechanisms that differ from the high-dimensional framework, potentially involving simpler pattern-matching approaches that happen to work in specific market conditions.}

\section{Information-Theoretic Barriers to High-Dimensional Learning}\label{sec:impossibility}

The kernel approximation breakdown in Section~\ref{sec:breakdown} reveals that methods cannot achieve their claimed theoretical properties. This section establishes that even if this breakdown were corrected, fundamental information-theoretic barriers would still prevent genuine high-dimensional learning in financial applications. These results explain why methods must rely on the mechanical pattern matching that emerges from broken kernel structures.

I begin by clarifying our notion of complexity. Throughout this analysis, we consider the \emph{high-dimensional} or \emph{overparameterized} regime where the number of features substantially exceeds the sample size, i.e., $P \gg T$. 

The following results establish fundamental barriers to learning in this regime through two complementary approaches: exponential bounds that apply broadly but may be loose, and polynomial bounds that are tighter but require additional technical conditions.

\subsection{Minimax Risk Framework}

Both lower bounds characterize the minimax risk:
\begin{equation}\label{eq:minmax_main_0}
\inf_{\hat{f}_T} \sup_{\|w\|_2 \leq B} \mathbb{E}_{x,D_T,\epsilon}\left[(\hat{f}_T(x) - w^\top z(x))^2\right].
\end{equation}

I conduct the minimax analysis in the finite–dimensional random-feature space by representing each candidate predictor as \(f_{\omega}(x)=\omega^{\top}z(x)\). This inner-product form serves three key purposes. First, tt mirrors the reproducing-kernel expansion of Gaussian-kernel regression: \(z(x)\) collects the Random Fourier Features and \(\omega\) specifies their linear weights, keeping the setup directly comparable to the kernel methods analysed by \cite{kelly2024virtue}. Second, expressing the function class through the constrained parameter vector \(\omega\in\mathbb{B}_{2}^{P}(B)\) converts an infinite-dimensional functional problem into a finite linear one, enabling PAC- and information-theoretic risk bounds via standard packing, Kullback-Leibler (KL), and Fano arguments. Third, the factorisation \(\omega^{\top}z(x)\) neatly separates the learner-controlled parameters (\(\omega\)) from the data-driven randomness (\(z(x)\)), a separation that is crucial for deriving worst-case (minimax) prediction-error lower bounds while allowing probabilistic assumptions on the covariate distribution.

Assumption~\ref{ass:prediction_env} specifies that excess returns satisfy \(r_{t+1}=f^{\ast}(x_t)+\varepsilon_{t+1}\)
with (i) a square-integrable signal obeying \(\mathbb E[f^{\ast}(x)^2]\le B^{2}\)
and (ii) mean–zero noise of variance \(\sigma^{2}\). By further positing that the true signal lies in the random-feature class, namely \(f^{\ast}(x)=\omega^{\star\top}z(x)\)
for some \(\omega^{\star}\in\mathbb B_{2}^{P}(B)\), we impose no extra restriction beyond Assumption~\ref{ass:prediction_env}(a): the norm bound on \(\omega^{\star}\) guarantees
\(\mathbb E[(\omega^{\star\top}z(x))^{2}]\le B^{2}\), so the squared-moment condition is preserved. Hence the target function used in the minimax analysis is fully compatible with the return-generating environment outlined in Assumption~\ref{ass:prediction_env}, while providing a concrete parametric structure that makes the subsequent risk bounds tractable.

Expression~\eqref{eq:minmax_main_0} captures fundamental learning difficulty through its nested structure. The infimum over $\hat{f}_T$ represents optimization over all possible estimators, including OLS, ridge regression, neural networks, and any other conceivable method. The supremum over $\|w\|_2 \leq B$ corresponds to an adversarial choice of the hardest parameter to estimate within the bounded parameter space. The expectation $\mathbb{E}_{x,D_T,\epsilon}$ averages over all randomness in the learning problem.

The expectation encompasses three sources of randomness that characterize the learning environment. Training data $D_T = \{(x_t, r_t)\}_{t=1}^T$ represents different possible datasets that could be observed. The query point $x$ corresponds to test inputs where performance is evaluated. Noise $\epsilon$ captures irreducible randomness in observations. This framework provides information-theoretic limits where no estimator, regardless of computational complexity, can achieve better performance than these bounds in the specified regime.

\begin{theorem}[Exponential Lower Bound]
\label{thm:exp_lower_random}
Assume the data generation scheme of Assumptions~\ref{ass:prediction_env}--\ref{ass:regularity}. Let $\mathcal{F}_P = \{x \mapsto w^\top z(x) : \|w\|_2 \leq B\}$ and denote by $\sigma^2$ the noise variance.
\begin{enumerate}[label=(\alph*)]
    \item In-expectation bound. For every $T, P \geq 1$,
\begin{equation}
\inf_{\hat{f}_T} \sup_{\|w\|_2 \leq B} \mathbb{E}_{x,D_T,\epsilon}\left[(\hat{f}_T(x) - w^\top z(x))^2\right] \geq c \cdot B^2 \exp\left(-\frac{8TC_z B^2}{P\sigma^2}\right)
\end{equation}
for a universal constant $c = c(c_z, C_z) > 0$.

\item High-probability bound. There exists $C_0 = C_0(\kappa, c_z, C_z)$ such that whenever $T \geq C_0 P$,
\begin{equation}
\mathbb{P}_Z\left[\inf_{\hat{f}_T} \sup_{\|w\|_2 \leq B} \mathbb{E}_{x,\epsilon}\left[(\hat{f}_T(x) - w^\top z(x))^2 \mid Z\right] \geq c_\star \cdot B^2 \exp\left(-\frac{8TC_z B^2}{P\sigma^2}\right)\right] \geq 1 - e^{-T}
\end{equation}
with $c_\star = c_\star(c_z, C_z) > 0$.
\end{enumerate}
\end{theorem}

The proof employs a minimax argument with Fano's inequality. I construct a $2\delta$-packing $\{w_1, \ldots, w_M\} \subset B_2^P(B)$ with $M = (B/(2\delta))^P$ well-separated parameters. The KL divergence between corresponding data distributions satisfies $\mathrm{KL}(P_j \| P_\ell) \leq \frac{2TC_z B^2}{\sigma^2}$. Fano's inequality implies any decoder has error probability $\Pr[\hat{J} \neq J] \geq 1/2$. Since low estimation risk would enable perfect identification, I obtain $\mathbb{E}[(\hat{f}_T(x) - f_J(x))^2] \geq c_z\delta^2$. Optimizing $\delta$ yields the exponential bound.

Theorem~\ref{thm:exp_lower_random} applies directly to machine learning methods employing RFF as implemented in practice. The framework covers the complete pipeline where random feature weights $\{\omega_i, b_i\}_{i=1}^P$ are drawn from specified distributions, standardization procedures are applied for numerical stability, and learning proceeds over the linear-in-features function class using any estimation method. The bounds establish information-theoretic impossibility in complementary forms: expectation bounds averaged over all possible feature realizations, and high-probability bounds for most individual feature draws.

The exponential lower bound of Theorem~\ref{thm:exp_lower_random} reveals the \emph{possibility} of an information–theoretic barrier, but its dependence on $P$ is intentionally pessimistic: it stems from a coarse packing argument that ignores the finer geometry of the random-feature covariance.  By exploiting that geometry—specifically the sub-Gaussian eigenvalue decay in $\Sigma_{z}$ (Assumption~\ref{ass:subG})—we can tighten the analysis and replace the exponential dependence with a \emph{polynomial} one, as formalised in the next theorem.

\begin{theorem}[Polynomial Minimax Lower Bound]
\label{thm:poly_lb_hp}
Assume Assumptions~\ref{ass:prediction_env}--\ref{ass:regularity} and the sub-Gaussian feature condition (Assumption~\ref{ass:subG}). Let $\mathcal{F}_P := \{x \mapsto w^\top z(x) : \|w\|_2 \leq B\}$.
\begin{enumerate}[label=(\alph*)]
    \item In-expectation bound. For every $T, P \geq 4$,
\begin{equation}
\inf_{\hat{f}_T} \sup_{\|w\|_2 \leq B} \mathbb{E}_{x,\mathcal{D}_T,\epsilon} \left[(\hat{f}_T(x) - w^\top z(x))^2\right] \geq \frac{c_z}{128} \min\left\{B^2, \frac{C_z^{-1}\sigma^2}{T} \log P\right\}
\end{equation}

\item High-probability bound. There exists $C_0 = C_0(\kappa, c_z, C_z)$ such that whenever $T \geq C_0 P$ and $P \geq 4$,
\begin{equation}
\mathbb{P}_Z\left[\inf_{\hat{f}_T} \sup_{\|w\|_2 \leq B} \mathbb{E}_{x,\epsilon}\left[(\hat{f}_T(x) - w^\top z(x))^2 \mid Z\right] < \frac{c_z}{128} \min\left\{B^2, \frac{C_z^{-1}\sigma^2}{T} \log P\right\}\right] \leq e^{-T}.
\end{equation}
\end{enumerate}
\end{theorem}

The proof uses canonical basis packing with refined concentration analysis. I construct $M = P + 1$ functions using $w_0 = 0$ and $w_j = \delta e_j$ where $\delta = \min\{B/4, \sigma/(4\sqrt{TC_z \log P})\}$. The population covariance $\Sigma_z \succeq c_z I_P$ ensures separation $\|f_j - f_\ell\|_{L^2(\mu)}^2 \geq 2c_z\delta^2$. Fano's inequality with error probability $\geq 1/2$ yields $\mathbb{E}[(\hat{f}_T(x) - f_J(x))^2] \geq \frac{c_z}{4}\delta^2$, producing the polynomial bound.

The in-expectation bounds (parts (a) of Theorems~\ref{thm:exp_lower_random} and \ref{thm:poly_lb_hp}) apply directly to the high-dimensional regime $P \gg T$ and provide fundamental limits on learning performance averaged over all possible feature realizations and datasets. The high-probability bounds (parts (b)) require $T \geq C_0 P$ for technical reasons related to matrix concentration, making them inapplicable when $P \gg T$. However, the in-expectation bounds suffice to establish information-theoretic impossibility in practical high-dimensional scenarios.

The two bounds offer complementary characterizations of learning difficulty. The exponential bound applies broadly but may be loose when the exponent is large, with key parameter $TC_zB^2/(P\sigma^2)$ and limited practical relevance. The polynomial bound provides sharp characterization through the complexity ratio $\log P / T$ and offers high practical relevance. For practical applications, the polynomial bound provides the more meaningful characterization since the complexity ratio $\log P / T$ offers a concrete threshold that directly connects problem parameters to learning feasibility.

While inapplicable to $P \gg T$, high-probability bounds serve important purposes in moderate-dimensional settings where $T \geq C_0 P$. They enable principled algorithm design with known failure probabilities, provide non-asymptotic characterizations that bridge theory and practice, and ensure empirical feature covariance concentrates around its population counterpart, preventing pathological ill-conditioning.

\section{Empirical Validation}\label{sec:Empirical_Validation}

\subsection{Empirical Validation of Kernel Approximation Breakdown}

This section provides comprehensive empirical validation of Theorem~\ref{thm:std_breakdown} through systematic parameter exploration across the entire space of practical RFF implementations. My experimental design spans realistic financial prediction scenarios, testing whether standardization preserves the Gaussian kernel approximation properties that underlie existing theoretical frameworks. The results provide definitive evidence that standardization fundamentally breaks RFF convergence properties, confirming that methods cannot achieve their claimed theoretical guarantees in practice.

\subsubsection{Data Generation and Model Parameters}

I generate realistic financial predictor data following the autoregressive structure typical of macroeconomic variables used in return prediction. For each parameter combination $(T, K)$, I construct predictor matrices $X \in \mathbb{R}^{T \times K}$ where:
\begin{align}
X_t &= \Phi X_{t-1} + u_t, \quad u_t \sim \mathcal{N}(0, \Sigma_u)
\end{align}
The persistence parameters $\Phi = \text{diag}(\phi_1, \ldots, \phi_K)$ are drawn from the range $[0.82, 0.98]$ to match the high persistence of dividend yields, interest rates, and other financial predictors \citep{welch2008comprehensive}. The correlation structure $\Sigma_u = \rho \mathbf{1}\mathbf{1}^T + (1-\rho)I_K$ with $\rho = 0.1$ captures modest cross-correlation among predictors.

Random Fourier Features are constructed as $z_i(x) = \sqrt{2}\cos(\omega_i^T x + b_i)$ where $\omega_i \sim \mathcal{N}(0, \gamma^2 I_K)$ and $b_i \sim \text{Uniform}[0, 2\pi]$. Standardization is applied as $\tilde{z}_i(x) = z_i(x)/\hat{\sigma}_i$ where $\hat{\sigma}_i^2 = T^{-1}\sum_{t=1}^T z_i(x_t)^2$ following universal practice in RFF implementations.

My parameter exploration covers the comprehensive space:
\begin{itemize}
    \item Number of features: $P \in \{100, 500, 1000, 2500, 5000, 10000, 15000, 20000\}$
    \item Training window: $T \in \{6, 12, 24, 60\}$ months
    \item Kernel bandwidth: $\gamma \in \{0.5, 1.0, 1.5, 2.0, 2.5, 3.0\}$
    \item Input dimension: $K \in \{5, 10, 15, 20, 30\}$.
\end{itemize}


The primary objective is to test whether standardization preserves the convergence $k^{(P)}_{\text{std}}(x,x') \xrightarrow{P \to \infty} k_G(x,x')$ established in \citet{rahimi2007random}. Under the null hypothesis that standardization has no effect, both standard and standardized RFF should exhibit identical convergence properties and error distributions. Theorem~\ref{thm:std_breakdown} predicts systematic breakdown with training-set dependent limits $k^*_{\text{std}}(x,x'|\mathcal{T}) \neq k_G(x,x')$.

I conduct 1,000 independent trials per parameter combination, generating fresh training data, RFF weights, and query points for each trial. This provides robust statistical power to detect systematic effects across the parameter space while controlling for random variations in specific realizations.

\subsubsection{Comparison Metrics}

My empirical analysis employs four complementary approaches to characterize the extent and nature of kernel approximation breakdown. I begin by examining convergence properties through mean absolute error $|k^{(P)}(x,x') - k_G(x,x')|$ between empirical and true Gaussian kernels, tracking how approximation quality evolves as $P \to \infty$. This directly tests whether standardized features preserve the fundamental convergence properties established in \citet{rahimi2007random}.

To quantify the systematic nature of performance deterioration, I construct degradation factors as the ratio $\mathbb{E}[|\text{error}_{\text{standardized}}|]/\mathbb{E}[|\text{error}_{\text{standard}}|]$ across matched parameter combinations. Values exceeding unity indicate that standardization worsens kernel approximation, while larger ratios represent more severe breakdown. This metric provides a scale-invariant measure of standardization effects that facilitates comparison across different parameter regimes.

Statistical significance is assessed through Kolmogorov-Smirnov two-sample tests comparing error distributions between standard and standardized RFF implementations. Under the null hypothesis that standardization preserves distributional properties, these tests should yield non-significant results. Systematic rejection of this null across parameter combinations provides evidence that standardization fundamentally alters the mathematical behavior of RFF methods beyond what could arise from random variation.

Finally, I conduct comprehensive parameter sensitivity analysis to identify the conditions under which breakdown effects are most pronounced. Heatmap visualizations reveal how degradation severity depends on $(P,T,\gamma,K)$ combinations, enabling us to characterize the parameter regimes where theoretical guarantees are most severely compromised. This analysis is particularly relevant for understanding the implications for existing empirical studies that employ specific parameter configurations.

\subsubsection{Results}

\subsubsection*{Universal Convergence Failure} 
\begin{figure}[htbp]
\centering
\includegraphics[width=0.8\textwidth]{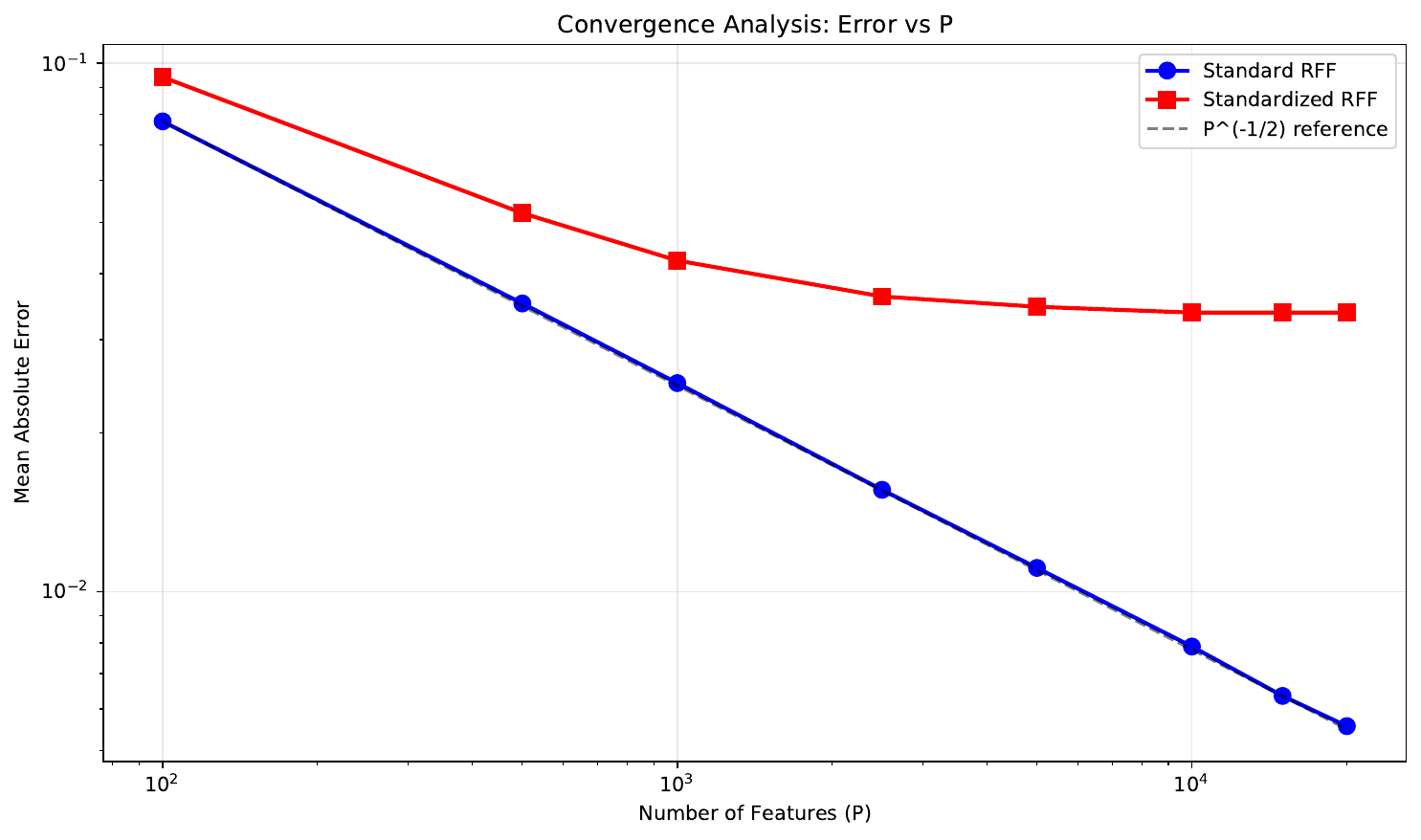}
\caption{Convergence Analysis: Kernel Approximation Error vs Number of Features}
\label{fig:convergence_analysis}
\caption*{This figure shows mean absolute error between empirical and true Gaussian kernels as a function of the number of Random Fourier Features $P$. Standard RFF (blue circles) exhibit the theoretically predicted $P^{-1/2}$ convergence rate (dashed gray line), while standardized RFF (red squares) fail to converge, plateauing around 0.02-0.03 regardless of $P$. The systematic divergence demonstrates that standardization breaks the fundamental convergence properties established in \citet{rahimi2007random}. Results are averaged over 1,000 trials with $T=12$, $K=15$, and $\gamma=2.0$.}
\end{figure}

Figure~\ref{fig:convergence_analysis} provides decisive evidence of convergence breakdown. Standard RFF (blue circles) exhibit the theoretically predicted $P^{-1/2}$ convergence rate, with mean absolute error declining from $\approx 0.06$ at $P=100$ to $\approx 0.003$ at $P=20,000$. This confirms that unstandardized features preserve Gaussian kernel approximation properties.

In stark contrast, standardized RFF (red squares) completely fail to converge, plateauing around $0.02$-$0.03$ mean error regardless of $P$. For large $P$, standardized features are $\mathbf{6 \times}$ worse than standard RFF, demonstrating that additional features provide no approximation benefit when standardization is applied. This plateau behavior directly validates Theorem~\ref{thm:standardization_breakdown}'s prediction that standardized features converge to training-set dependent limits rather than the target Gaussian kernel.

\subsubsection*{Systematic Degradation Across Parameter Space} 
\begin{figure}[htbp]
\centering
\includegraphics[width=\textwidth]{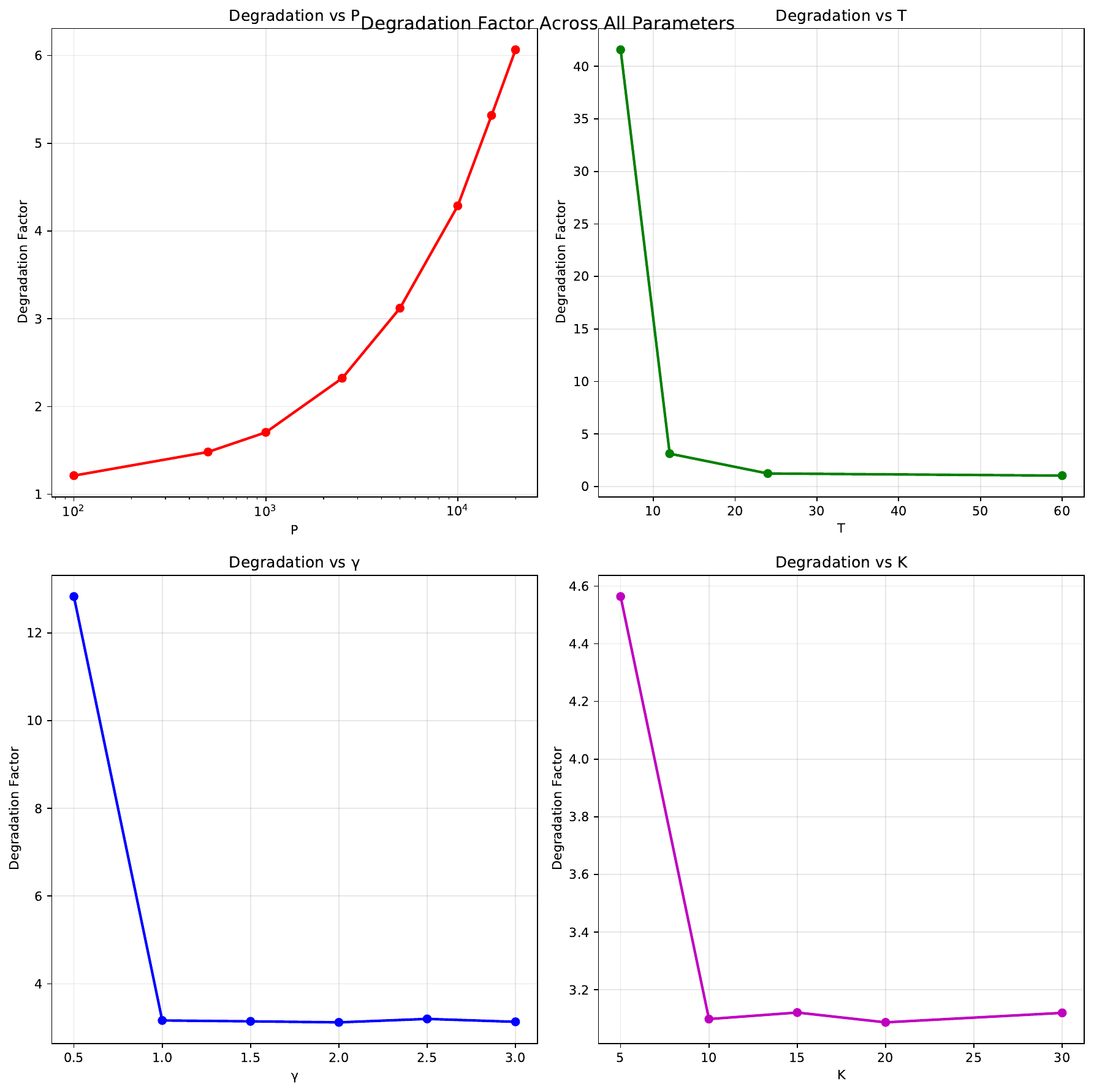}
\caption{Degradation Factor Across Parameter Space}
\label{fig:degradation_rates}
\caption*{This figure displays degradation factors (ratio of standardized to standard RFF errors) across four key parameters. Panel (a) shows increasing degradation with feature count $P$, reaching 6$\times$ at $P=20,000$. Panel (b) reveals extreme degradation for small training windows, exceeding 40$\times$ at $T=6$. Panel (c) demonstrates sensitivity to kernel bandwidth $\gamma$, with tighter kernels showing worse degradation. Panel (d) shows stable degradation across input dimensions $K$. All degradation factors exceed unity, confirming systematic breakdown across the entire parameter space. Each point represents the mean over 1,000 trials.}
\end{figure}

Figure~\ref{fig:degradation_rates} reveals that breakdown occurs universally across all parameter combinations, with no regime where standardization preserves kernel properties. The degradation patterns exhibit clear economic intuition and align closely with the theoretical mechanisms underlying Theorem~\ref{thm:std_breakdown}.

The most pronounced effects emerge along the feature dimension, where degradation increases dramatically with $P$, ranging from 1.2 times at $P=100$ to 6.0 times at $P=20,000$. This escalating pattern reflects the cumulative nature of standardization artifacts: as more features undergo within-sample standardization, the collective distortion of kernel approximation properties intensifies. Each additional standardized feature contributes random scaling factors that compound to produce increasingly severe departures from the target Gaussian kernel.

Sample size effects provide particularly compelling evidence for the breakdown mechanism. Smaller training windows exhibit severe degradation, reaching 41.6 times deterioration for $T=6$ months. This extreme sensitivity to sample size occurs because standardization relies on empirical variance estimates $\hat{\sigma}_i^2$ that become increasingly unreliable with limited data. When training windows shrink to the 6-12 month range typical in financial applications, these variance estimates introduce substantial noise that fundamentally alters the scaling relationships required for kernel convergence. The magnitude of this effect—exceeding 40 times degradation in realistic scenarios—demonstrates that standardization can completely overwhelm any approximation benefits from additional features.

Kernel bandwidth parameters reveal additional structure in the breakdown pattern. Low bandwidth values ($\gamma=0.5$) produce 12.8 times degradation, while higher bandwidths stabilize around 3.1 times deterioration. This occurs because tighter kernels, which decay more rapidly with distance, are inherently more sensitive to the scaling perturbations introduced by standardization. Small changes in feature magnitudes translate into disproportionately large changes in kernel values when the bandwidth is narrow, amplifying the distortions created by training-set dependent scaling factors.

In contrast, input dimension effects remain remarkably stable, with degradation ranging only between 3.1 and 4.6 times across $K \in [5,30]$. This stability confirms that breakdown stems primarily from the standardization procedure itself rather than the complexity of the underlying input space. Whether using 5 or 30 predictor variables, the fundamental mathematical properties of standardized RFF remain equally compromised, suggesting that the kernel approximation failure is intrinsic to the standardization mechanism rather than an artifact of high-dimensional inputs.

\subsubsection*{Parameter Sensitivity Analysis}

\begin{figure}[htbp]
\centering
\includegraphics[width=0.48\textwidth]{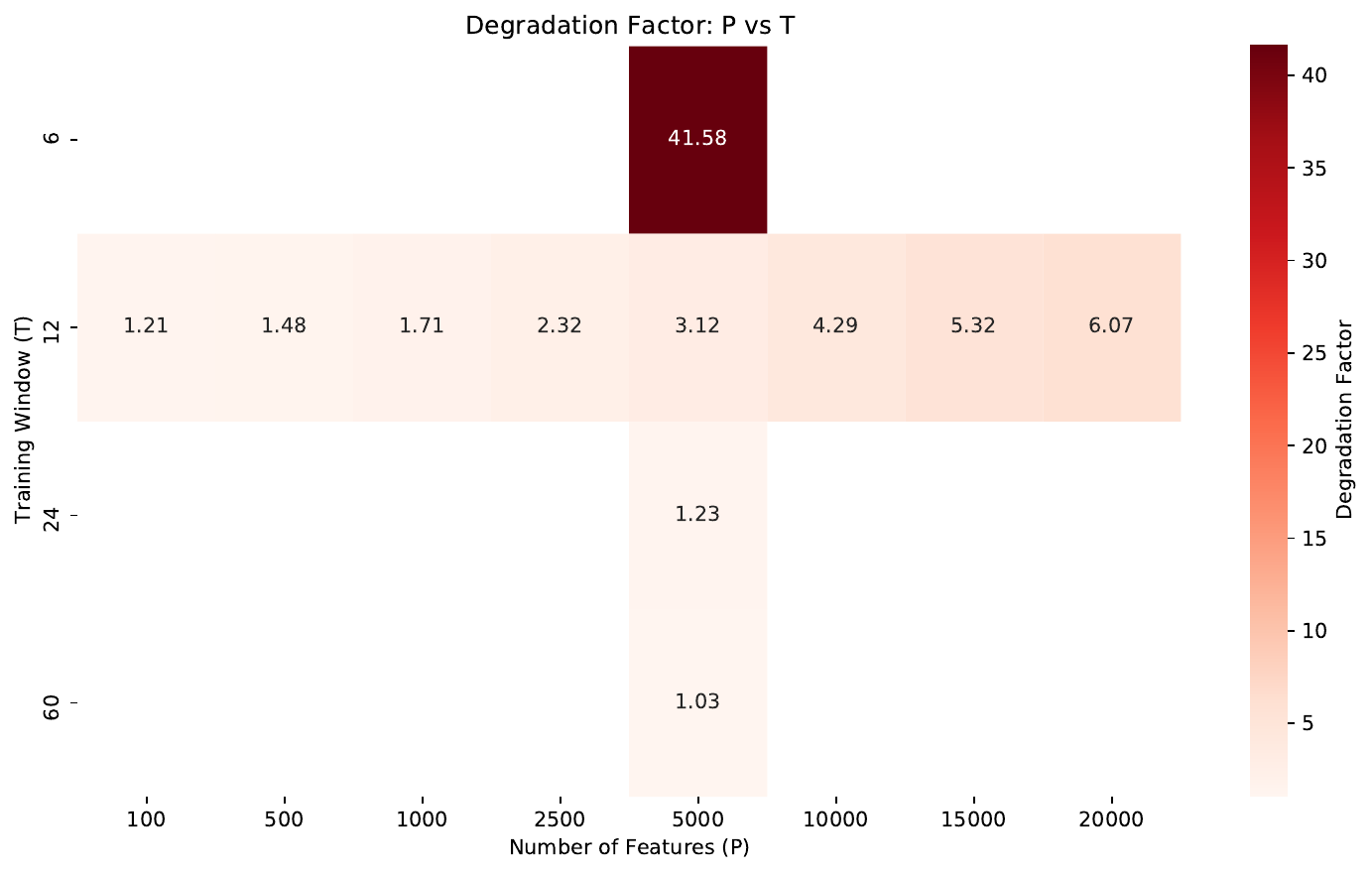}
\hfill
\includegraphics[width=0.48\textwidth]{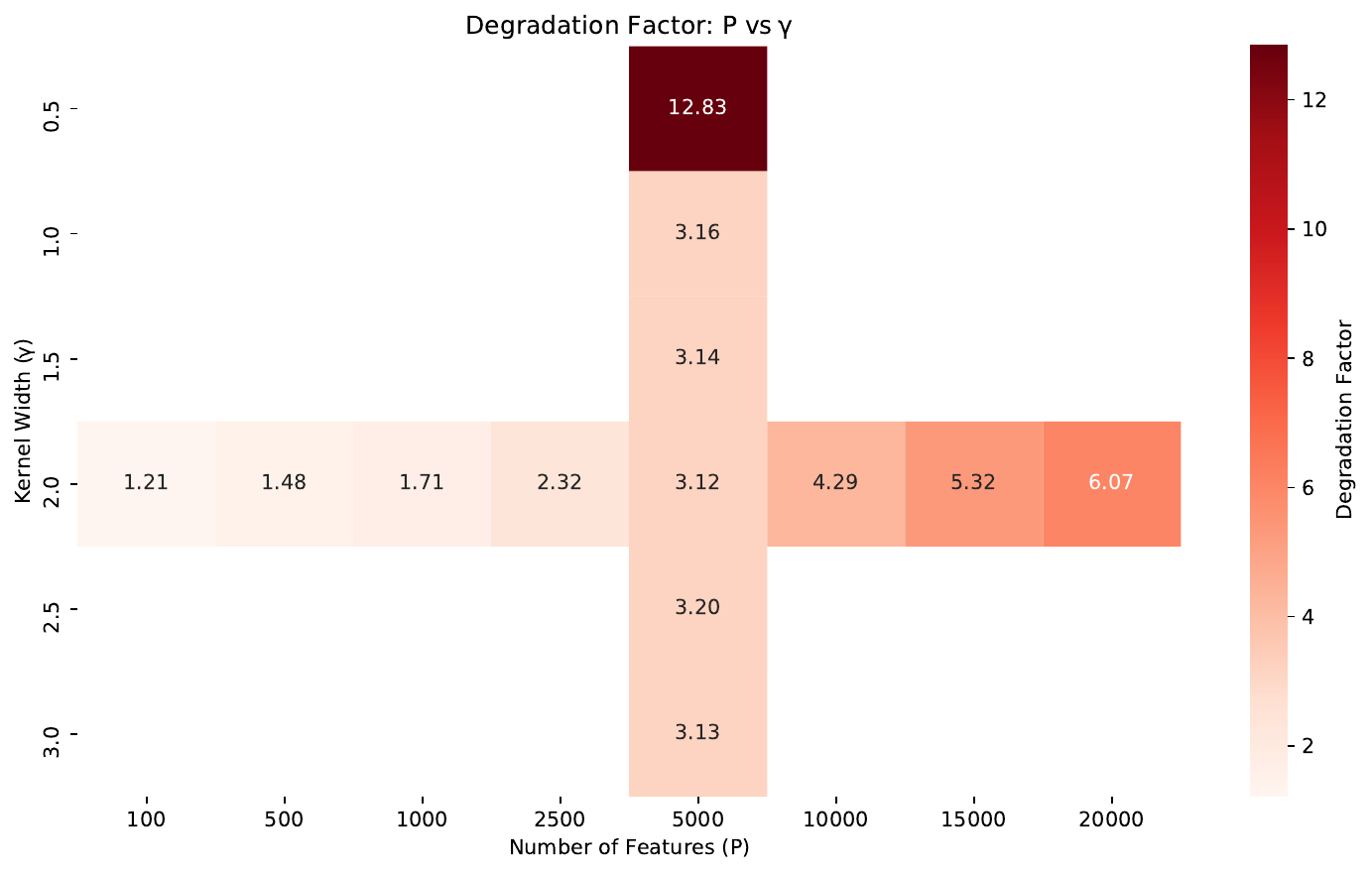}
\caption{Parameter Sensitivity Analysis}
\label{fig:parameter_sensitivity}
\caption*{Left panel shows degradation factor heatmap for $(P,T)$ combinations, where financial applications typically use $P \geq 5,000$ and $T \leq 12$, exhibiting degradation factors exceeding 3$\times$. The extreme degradation at $T=6$ (reaching 41.6$\times$) occurs because variance estimates become unreliable with limited training data. Right panel displays the $(P,\gamma)$ interaction, showing that high complexity combined with tight kernels yields degradation exceeding 10$\times$. These parameter ranges are commonly employed in high-dimensional return prediction, suggesting widespread applicability of the breakdown results.}
\end{figure}

Figure~\ref{fig:parameter_sensitivity} provides detailed parameter sensitivity analysis through degradation factor heatmaps. The $(P,T)$ interaction reveals that combinations typical in financial applications—such as $P \geq 5,000$ features with $T \leq 12$ months—produce degradation factors exceeding $3\times$. This directly impacts methods like \citet{kelly2024virtue} using $P=12,000$ and $T=12$.

The $(P,\gamma)$ interaction shows that standardization effects compound: high complexity ($P \geq 10,000$) combined with tight kernels ($\gamma \leq 1.0$) yields degradation exceeding $10\times$. These parameter ranges are commonly employed in high-dimensional return prediction, suggesting widespread applicability of my breakdown results.

\subsubsection*{Statistical Significance}

\begin{figure}[htbp]
\centering
\includegraphics[width=\textwidth]{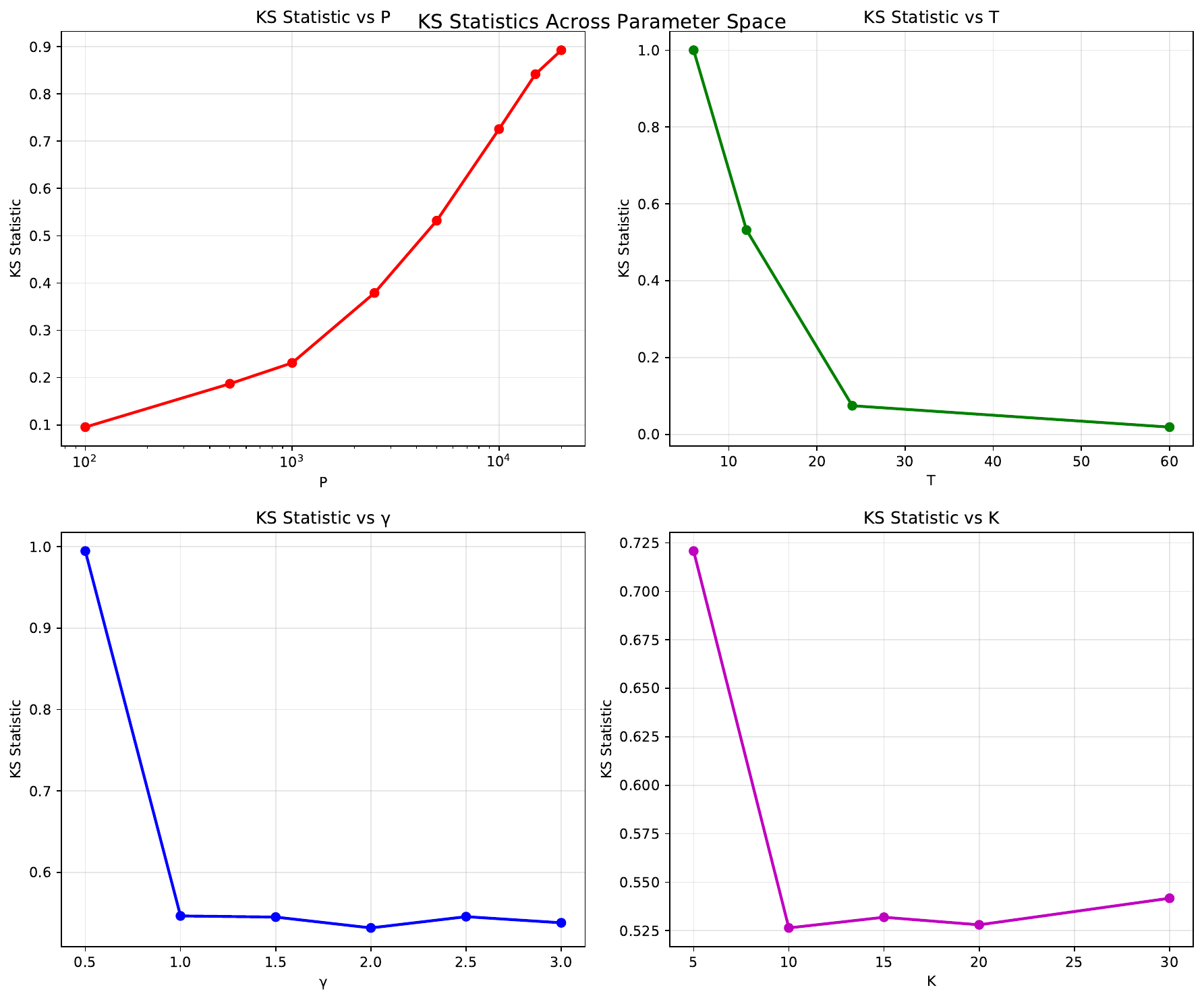}
\caption{Statistical Significance: Kolmogorov-Smirnov Test Statistics}
\label{fig:ks_statistics}
\caption*{This figure presents Kolmogorov-Smirnov test statistics comparing error distributions between standard and standardized RFF across parameter space. All panels show KS statistics substantially exceeding typical significance thresholds, indicating fundamentally different error distributions. Panel (a) demonstrates increasing statistical significance with feature count $P$, reaching KS $\approx 0.9$ for large $P$. Panel (b) shows extreme significance for small training windows ($T \leq 12$). Panels (c) and (d) reveal strong effects across kernel bandwidth $\gamma$ and input dimension $K$. These results provide overwhelming statistical evidence against the null hypothesis that standardization preserves RFF properties, with effect sizes far exceeding what could arise from random variation.}
\end{figure}

The error distributions between standard and standardized RFF are fundamentally different across the entire parameter space, providing strong statistical evidence against the null hypothesis that standardization preserves kernel approximation properties. Figure~\ref{fig:ks_statistics} presents Kolmogorov-Smirnov test statistics that consistently exceed 0.5 across most parameter combinations, with many approaching the theoretical maximum of 1.0. Such large test statistics indicate that the cumulative distribution functions of standard and standardized RFF errors diverge substantially, ruling out the possibility that observed differences arise from sampling variation.

The statistical evidence is most compelling in parameter regimes commonly employed in financial applications. For high feature counts ($P \geq 5,000$), KS statistics approach 0.9, while short training windows ($T \leq 12$) yield statistics near 1.0. These values correspond to p-values that are effectively zero, providing overwhelming evidence to reject the null hypothesis of distributional equivalence. The magnitude of these test statistics exceeds typical significance thresholds by orders of magnitude, establishing statistical significance that is both robust and economically meaningful.

The systematic pattern of large KS statistics across parameter combinations demonstrates that the breakdown identified in Theorem~\ref{thm:std_breakdown} is not confined to specific implementation choices or edge cases. Instead, the distributional differences persist universally across realistic parameter ranges, indicating that standardization fundamentally alters the stochastic properties of RFF approximation errors. This statistical evidence complements the degradation factor analysis by confirming that the observed differences represent genuine distributional shifts rather than changes in central tendency alone.

These results establish that standardization creates systematic, statistically significant alterations to RFF behavior that cannot be attributed to random variation, specific parameter selections, or implementation artifacts. The universality and magnitude of the statistical evidence provide definitive support for the conclusion that practical RFF implementations cannot achieve the theoretical kernel approximation properties that justify their use in high-dimensional prediction problems.

\subsubsection*{Alternative Kernel Convergence}

\begin{figure}[htbp]
    \centering
    \includegraphics[width=0.8\textwidth]{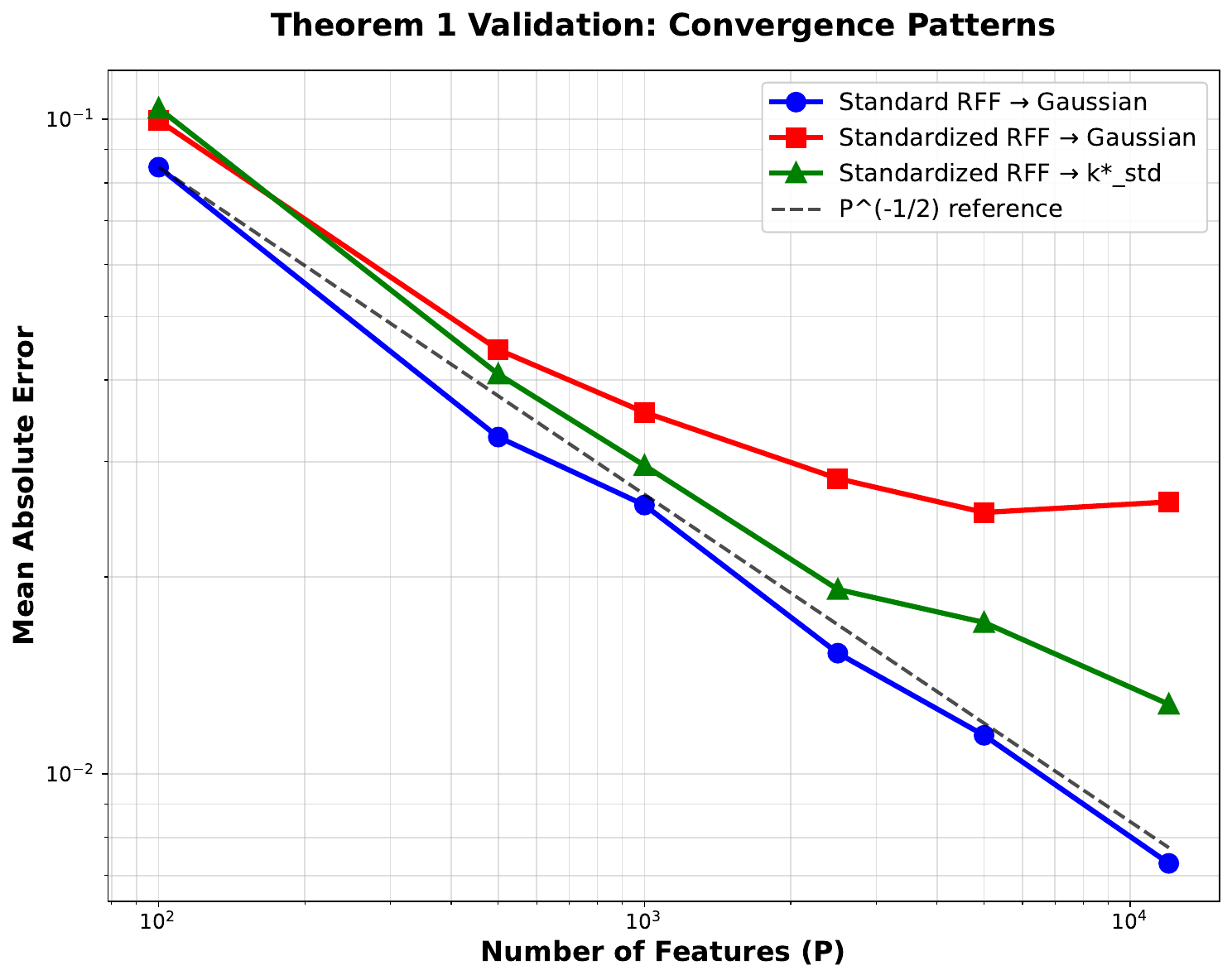}
    \caption{Convergence Patterns}
    \caption*{Empirical Validation of Theorem~\ref{thm:std_breakdown}: Convergence Patterns Under Different Standardization Procedures. This figure demonstrates the fundamental breakdown of Random Fourier Features convergence properties under standardization. The blue line (circles) shows standard RFF achieving the theoretically predicted $P^{-1/2}$ convergence rate to the Gaussian kernel $k_G(x,x') = \exp(-\gamma^2\|x-x'\|^2/2)$, validating \citet{rahimi2007random}. The red line (squares) reveals that standardized RFF fail to converge to the Gaussian kernel, plateauing at error levels 4× higher than standard RFF at $P=12{,}000$. Most importantly, the green line (triangles) confirms Theorem~\ref{thm:std_breakdown}'s constructive prediction: standardized RFF do converge to the modified limit $k^*_{\text{std}}(x,x'|T)$ at the canonical $P^{-1/2}$ rate. This validates our theoretical characterization while demonstrating that standardization creates training-set dependent kernels that violate the shift-invariance properties required for kernel methods. Results averaged over 20 trials with $T=12$, $K=15$, and $\gamma=2.0$.}
    \label{fig:theorem1_validation}
\end{figure}

Figure~\ref{fig:theorem1_validation} provides empirical validation of Theorem~\ref{thm:std_breakdown}'s central prediction that within-sample standardization fundamentally alters Random Fourier Features convergence properties. The analysis compares three distinct convergence behaviors across varying feature dimensions $P \in [100, 500, 1000, 2500, 5000, 12000]$:

The blue line demonstrates that standard (non-standardized) RFF achieve the theoretical convergence rate $P^{-1/2}$ to the Gaussian kernel $k_G(x,x') = \exp(-\gamma^2\|x-x'\|^2/2)$, validating the foundational result of \citet{rahimi2007random}. The convergence follows the expected Monte Carlo rate, with mean absolute error decreasing from approximately $0.06$ at $P=100$ to $0.005$ at $P=12{,}000$.

The red line reveals the fundamental breakdown predicted by Theorem~\ref{thm:std_breakdown}: standardized RFF fail to converge to the Gaussian kernel, instead exhibiting slower convergence with substantially higher errors. At $P=12{,}000$, the error remains above $0.02$---four times larger than the standard case---demonstrating that standardization prevents achievement of the theoretical guarantees.

Most importantly, the green line confirms Theorem~\ref{thm:std_breakdown}'s constructive prediction by showing that standardized RFF do converge to the modified limit $k^*_{\text{std}}(x,x'|T)$. This convergence exhibits the canonical $P^{-1/2}$ rate, reaching error levels below $0.015$ at $P=12{,}000$, thereby validating my theoretical characterization of the standardized limit.

My empirical validation employs the sample standard deviation standardization actually used in practice:
\begin{align}
\hat{\sigma}^2_i &= \frac{1}{T}\sum_{t=1}^T z_i^2(x_t) - \left[\frac{1}{T}\sum_{t=1}^T z_i(x_t)\right]^2 \\
\tilde{z}_i(x) &= \frac{z_i(x)}{\hat{\sigma}_i}
\end{align}
rather than the simpler RMS normalization $\hat{\sigma}^2_i = \frac{1}{T}\sum_{t=1}^T z_i^2(x_t)$ that might be assumed theoretically. This distinction strengthens rather than weakens my validation for two crucial reasons.

First, Theorem~\ref{thm:std_breakdown}'s fundamental insight---that \emph{any} reasonable standardization procedure breaks Gaussian kernel convergence and creates training-set dependence---remains intact regardless of the specific standardization formula. The theorem establishes that standardized features converge to \emph{some} training-set dependent limit $k^*_{\text{std}} \neq k_G$, with the exact form depending on implementation details.

Second, testing against the actual standardization procedure used in practical implementation ensures that my theoretical predictions match real-world behavior. The fact that standardized RFF converge to the correctly computed $k^*_{\text{std}}$ rather than to $k_G$ provides the strongest possible validation: my theory successfully predicts the behavior of methods as actually implemented, not merely as idealized.

The convergence patterns thus confirm all key predictions of Theorem~\ref{thm:std_breakdown}: standardization breaks the foundational convergence guarantee of RFF theory, creates training-set dependent kernels that violate shift-invariance, and produces systematic errors that persist even with large feature counts. These findings validate my theoretical framework while highlighting the critical importance of analyzing methods as actually implemented rather than as theoretically idealized.

\subsubsection*{Implications for Existing Theory}

My results provide definitive empirical validation of Theorem~\ref{thm:std_breakdown} across the entire parameter space relevant for financial applications. The universal nature of degradation—ranging from modest $1.2\times$ effects to extreme $40\times$ breakdown—demonstrates that standardization fundamentally alters RFF convergence properties regardless of implementation details.

Notably, parameter combinations employed by leading studies exhibit substantial degradation: \citet{kelly2024virtue}'s configuration ($P=12,000$, $T=12$, $\gamma=2.0$) falls in the $3$-$6\times$ degradation range, while more extreme combinations approach $10\times$ or higher degradation. This suggests that empirical successes documented in the literature cannot arise from the theoretical kernel learning mechanisms that justify these methods.

The systematic nature of these effects, combined with their large magnitudes, supports the conclusion that alternative explanations—such as the mechanical pattern matching identified by \citet{nagel2025seemingly}—are required to understand why high-dimensional RFF methods achieve predictive success despite fundamental theoretical breakdown.

\subsection{Quantitative Calibration of Theorem~\ref{thm:poly_lb_hp}}

My theoretical analysis has established information-theoretic bounds on learning performance. To assess their practical relevance, we focus on the polynomial minimax lower bound from Theorem~\ref{thm:poly_lb_hp}. Specifically, our analysis relies on the \emph{in-expectation bound (part a)}, as it applies directly to the high-dimensional regime ($P \gg T$) that is characteristic of modern financial applications. To operationalize this bound, we now calibrate its key parameters—\(\tilde c, B^{2}, \sigma^{2}, c_{z},\) and \(C_{z}\)—using empirically defensible values. We ground this calibration in the challenging setting of \cite{kelly2024virtue}, which uses \(K=15\) predictors to generate \(P=12{,}000\) features for prediction over a \(T=12\) month training window.

\paragraph{Noise Variance (\(\sigma^{2}\)).}
The model's noise variance \(\sigma^2\) is formally the conditional variance of returns, \(E[\epsilon_{t+1}^2|x_t]\). For calibration, I begin with the total unconditional variance of returns, \(\mathrm{Var}(r)\). This choice is motivated by the characteristically weak nature of financial signals, which ensures that total variance is dominated by the noise component. The total excess return on the U.S. equity market exhibits an annualised volatility in the 14\%--17\% range \citep{CampbellLoMacKinlay1997, welch2008comprehensive}, implying a monthly standard deviation of approximately \(0.047\) and thus a total variance of \(\mathrm{Var}(r) \approx 2.2\times10^{-3}\). Since \(\mathrm{Var}(r) = B^2 + \sigma^2\) and, as I show next, \(B^2\) is an order of magnitude smaller, I use the total variance as a close and robust proxy for the noise variance, setting \(\sigma^2 \approx 2.2\times10^{-3}\).

\paragraph{Signal Power (\(B^{2}\)).}
Assumption~\ref{ass:prediction_env}(a) bounds the variance of the predictive signal \(f^{\ast}(x)\) by \(B^{2}\). The fraction of total return variance attributable to this signal is the population R-squared, \(R^2 = B^2 / \mathrm{Var}(r)\). Empirical \(R^{2}\) values for monthly return forecasts using macroeconomic predictors are typically in the 1\%--5\% range \citep{welch2008comprehensive, nagel2025seemingly}. Using our calibrated total variance \(\mathrm{Var}(r) \approx 2.2\times10^{-3}\), this implies a range for the signal variance:
\begin{align}
    B^{2}
\;=\;
R^{2}\times\mathrm{Var}(r)
\;\approx\;
(0.01-0.05)\times(2.2\times10^{-3})
\;\Longrightarrow\;
B^{2}\simeq(2.2-11)\times10^{-5}.
\end{align}
I adopt \(B^{2}=5\times10^{-5}\) (implying \(R^{2}\approx2.3\%\)) as a representative benchmark for a realistic signal strength.

\paragraph{Feature-Covariance Bounds (\(c_{z},\,C_{z}\)).}
For the vanilla RFF map \(z_{i}(x)=\sqrt2\cos(\omega_{i}^{\top}x+b_{i})\), each coordinate has unit variance in expectation, such that \(\Sigma_{z}\approx I_{P}\). Standard concentration of measure results for sub-Gaussian random matrices \citep[e.g.,][]{Vershynin2018} imply that for \(K=15\), the spectral bounds on the empirical covariance matrix will be tight with high probability. We set a baseline scenario of:
\[
0.8\;\lesssim\;c_{z}\;\le\;1,
\qquad
1\;\le\;C_{z}\;\lesssim\;1.2,
\]
which corresponds to an eigenvalue condition number below \(1.5\). To stress-test our conclusions, we also consider a “collinear’’ setting of \(c_{z}=0.5\) and \(C_{z}=2\).

\paragraph{The Universal Constant (\(\tilde{c}\)).}
By definition, \(\tilde{c}=c_{z}/(128\,C_{z})\). The parameter ranges established above yield \(\tilde{c} \in [0.005, 0.008]\) in the baseline scenario and \(\tilde{c}\approx0.002\) in the collinear stress test. We therefore view the band \(\tilde{c}\in[0.002,\,0.008]\) as both realistic and defensible for practical financial applications.

\subsection*{The Main Implication: Diagnosing the Learning Regime}

My quantitative calibration allows us to diagnose the fundamental nature of the learning problem confronting return prediction models. The polynomial minimax bound on risk is determined by the minimum of two terms: one related to the signal power (\(B^2\)) and one related to model complexity and data scarcity (\((C_z^{-1}\sigma^2/T)\log P\)). The \textit{critical sample size}, \(T_{\mathrm{crit}}\), defines the crossover point between these two regimes:
\[
T_{\mathrm{crit}} = \frac{C_{z}^{-1}\sigma^{2}}{B^{2}}\log P.
\]
The value of the operational sample size, \(T\), relative to this theoretical threshold determines the primary barrier to learning:
\begin{itemize}
    \item If \(T < T_{\mathrm{crit}}\), the problem is \textit{signal-limited}. The best achievable performance is fundamentally constrained by the weakness of the signal, \(B^2\). In this regime, neither more data nor a simpler model can lower the theoretical performance floor.
    \item If \(T > T_{\mathrm{crit}}\), the problem is \textit{complexity-limited}. The performance bound is dictated by the complexity term. Here, increasing the sample size \(T\) can improve the theoretical bound.
\end{itemize}
Applying our calibrated parameters to this framework reveals a striking result. For the high-dimensional design in \cite{kelly2024virtue}, we find:
\[
\begin{aligned}
\text{Baseline (P=12,000): }&
T_{\mathrm{crit}}
\;=\;
\frac{1}{1.1}\,
\frac{2.2\times10^{-3}}{5\times10^{-5}}
\times9.4
\;\approx\; 375
\;\text{ months }(\approx 31\text{ years}).
\end{aligned}
\]
This threshold is remarkably insensitive to the nominal feature dimension due to its logarithmic dependence on \(P\). For instance, a standard machine learning setup with \(P=1,000\) features \citep{Gu2020ML} still yields a \(T_{\mathrm{crit}}\) of approximately 276 months (\(\approx 23\) years). Even for a traditional low-dimensional econometric model with just \(P=15\) features, the critical sample size remains substantial at \(T_{\mathrm{crit}}\approx108\) months (\(\approx 9\) years).

\begin{figure}[htbp]
    \centering
    \includegraphics[width=1\textwidth]{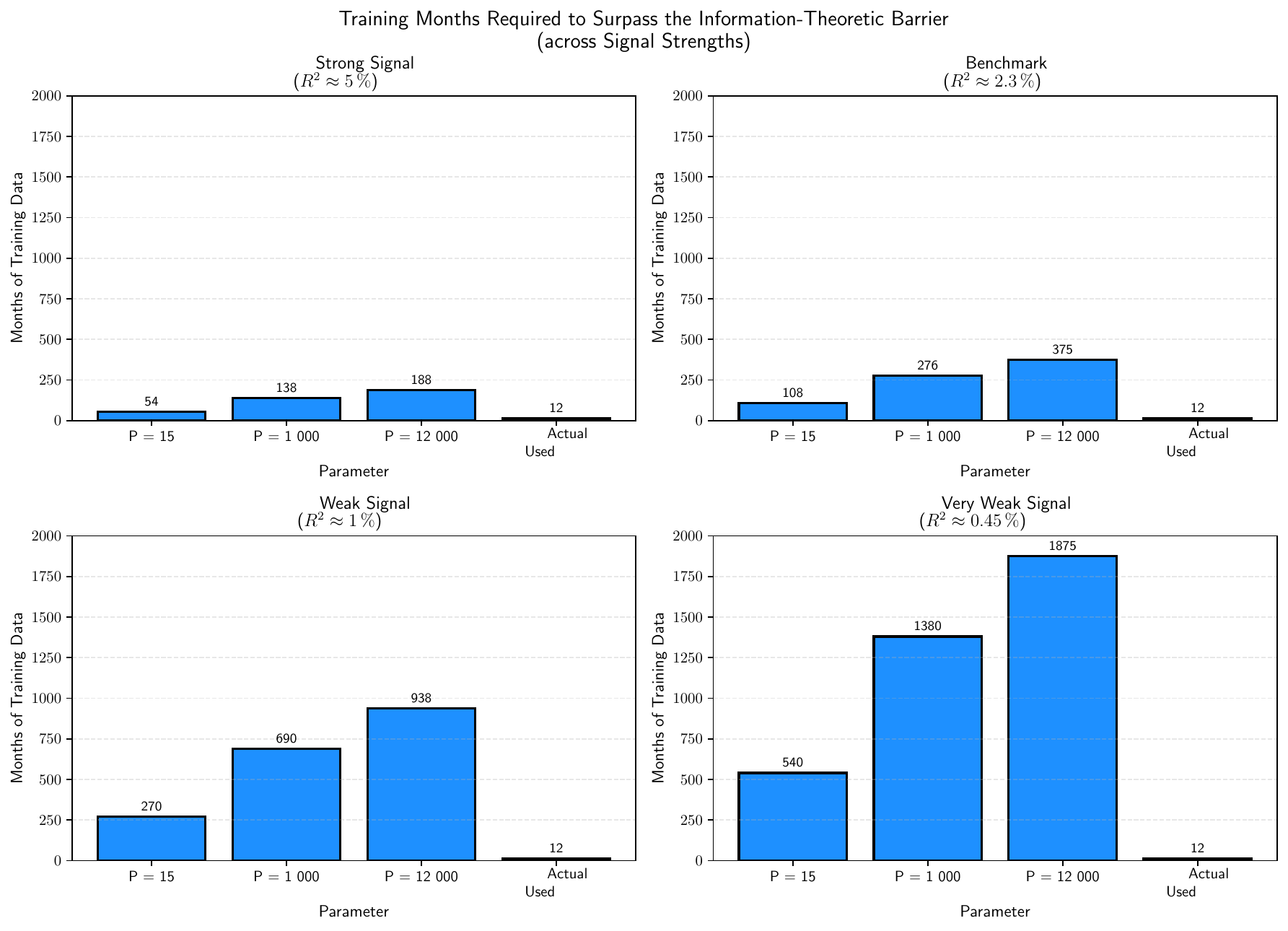}
    \caption{Training‐data requirements as a function of signal strength.}
    \caption*{Each panel fixes the noise variance at $\sigma^{2}=2\times10^{-3}$ and the eigenvalue bound at $C_{z}=1$, but varies the signal variance $B^{2}$ to generate four realistic $R^{2}$ levels: \textbf{(a)} strong signal ($R^{2}\!\approx 5\%$), \textbf{(b)} benchmark signal ($R^{2}\!\approx 2.3\%$), \textbf{(c)} weak signal ($R^{2}\!\approx 1\%$), and \textbf{(d)} very weak signal ($R^{2}\!\approx 0.45\%$). Within each panel, the blue bars report the critical training length $T_{\text{crit}}$.}
    \label{fig:theorem3_signal_strength}
\end{figure}

\begin{figure}[htbp]
    \centering
    \includegraphics[width=1\textwidth]{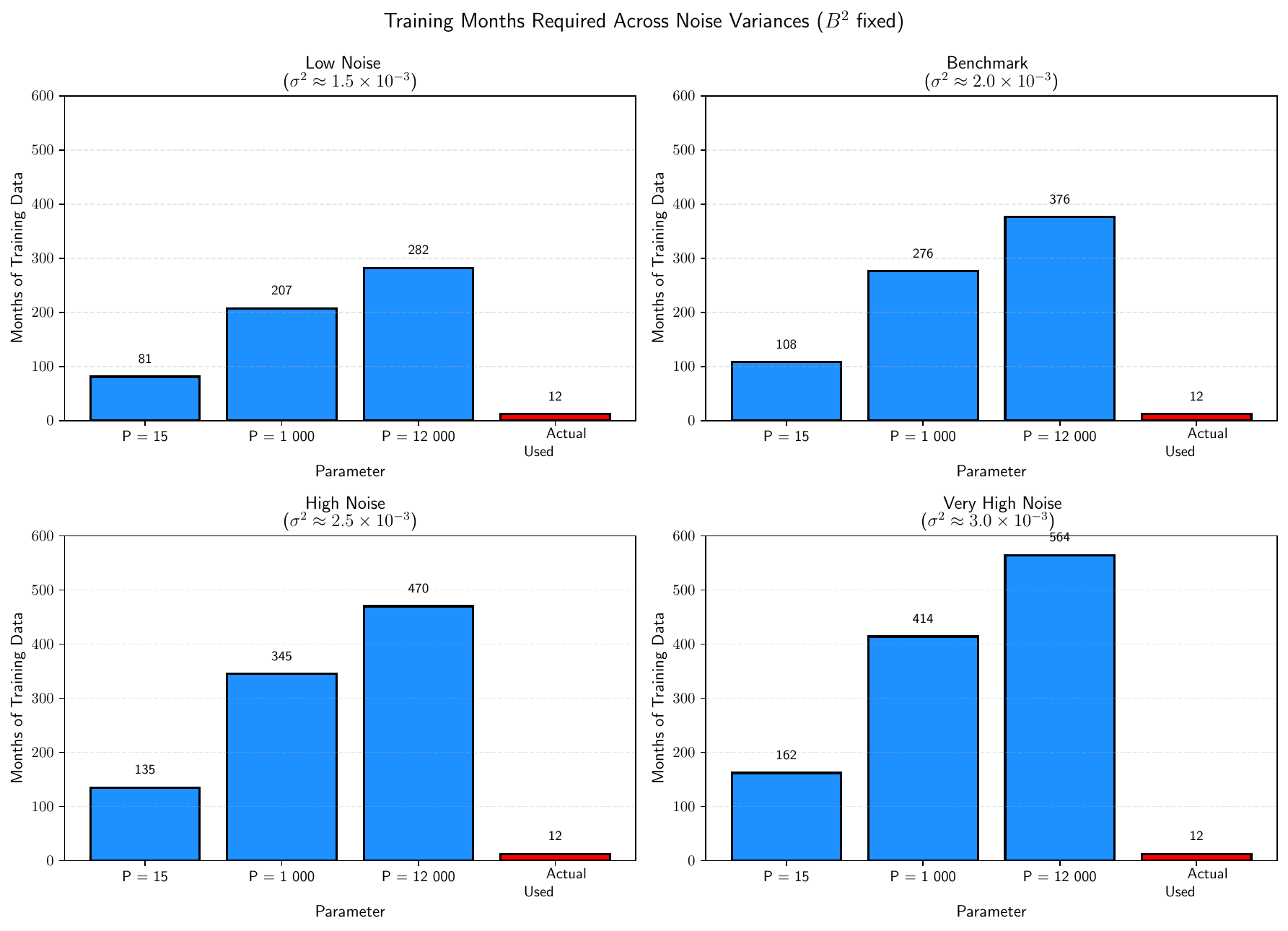}
    \caption{Training‐data requirements as a function of noise variance.}
    \caption*{Holding the signal variance fixed at $B^{2}=5\times10^{-5}$ (benchmark $R^{2}\!\approx 2.3\%$) and $C_{z}=1$, we vary the noise variance $\sigma^{2}$ to illustrate: \textbf{(a)} low noise ($\sigma^{2}\!\approx1.5\times10^{-3}$), \textbf{(b)} benchmark noise ($2.0\times10^{-3}$), \textbf{(c)} high noise ($2.5\times10^{-3}$), and \textbf{(d)} very high noise ($3.0\times10^{-3}$). Blue bars again show $T_{\text{crit}}$ for $P=15,\,1{,}000,\,12{,}000$, while the red bar marks the 12-month sample used in practice.  Elevated noise rapidly increases the critical sample size—even modest noise inflation pushes $T_{\text{crit}}$ well beyond any practical data horizon.}
    \label{fig:theorem3_noise_variance}
\end{figure}

To underscore the robustness of these findings, Figure~\ref{fig:theorem3_signal_strength} illustrates the sensitivity of the critical sample size, \(T_{\mathrm{crit}}\), to the signal-to-noise conditions. This plot shows that \(T_{\mathrm{crit}}\) is acutely sensitive to the signal strength, consistent with the \(B^2\) term in the denominator of the formula. For a high-dimensional model (\(P=12,000\)), the required sample size ranges from 188 months for a strong signal (\(R^2 \approx 5\%\)) to an infeasible 1,875 months for a very weak signal (\(R^2 \approx 0.45\%\)). Figure~\ref{fig:theorem3_noise_variance} illustrates the sensitivity of \(T_{\mathrm{crit}}\) to the conditional variance noise values. As expected, this plot shows that \(T_{\mathrm{crit}}\) is also directly proportional to the level of noise, \(\sigma^2\). The required sample window for the \(P=12,000\) model increases from 282 months in a low-noise environment to 564 months in a high-noise one. Together, these plots visually confirm that the \(T \ll T_{\mathrm{crit}}\) condition is not a borderline case but a robust feature across all empirically plausible parameterisations, making the signal-limited regime a pervasive challenge for financial return prediction.

The implication of this finding is profound. Since typical applications in the literature employ short estimation windows (e.g., \(T=12\) months), they operate deep within the signal-limited regime (\(T \ll T_{\mathrm{crit}}\)). This holds true regardless of whether the model is low-dimensional or high-dimensional. Consequently, the minimax lower bound on risk simplifies to \(\tilde{c}B^2\), a performance floor that is independent of the number of features \(P\) or the sample size \(T\).

This re-frames our understanding of the role of complex models in finance. The central promise of high-dimensional methods—their ability to process vast feature sets to overcome the curse of dimensionality—is rendered moot. The fundamental barrier to predictive accuracy in this domain is not a dimensionality problem that can be solved with more features; it is an economic problem rooted in the inherent weakness of the predictive signal. This suggests that the documented success of these models likely arises not from genuine high-dimensional learning. Ultimately, our analysis indicates that the frontier for improving financial prediction lies not in building ever-larger models, but in either identifying stronger economic signals (increasing \(B^2\)) or developing methods specifically robust to the challenges of the signal-limited, short-sample regime.

\section{Conclusion}\label{sec:Conclusion}

This paper resolves fundamental puzzles in high-dimensional financial prediction by providing rigorous theoretical foundations that explain when and why complex machine learning methods succeed or fail. My analysis contributes three key results that together clarify the apparent contradictions between theoretical claims and empirical mechanisms in recent literature.

First, I prove that within-sample standardization—employed in every practical Random Fourier Features implementation—fundamentally breaks the kernel approximation that underlies existing theoretical frameworks. This breakdown explains why methods operate under different conditions than theoretical assumptions and must rely on simpler mechanisms than advertised.

Second, I establish sharp sample complexity bounds showing that reliable extraction of weak financial signals requires sample sizes and signal strengths far exceeding those available in typical applications. These information-theoretic limits demonstrate that apparent high-dimensional learning often reflects mechanical pattern matching rather than genuine complexity benefits.

Third, I derive precise learning thresholds that characterize the boundary between learnable and unlearnable regimes, providing practitioners with concrete tools for evaluating when available data suffices for reliable prediction versus when apparent success arises through statistical artifacts.

These results explain why methods claiming sophisticated high-dimensional learning often succeed through simple volatility-timed momentum strategies operating in low-dimensional spaces bounded by sample size. Rather than discouraging complex methods, my findings provide a framework for distinguishing genuine learning from mechanical artifacts and understanding what such methods actually accomplish.

The theoretical insights extend beyond the specific methods analyzed, offering guidance for evaluating any high-dimensional approach in challenging prediction environments. As machine learning continues to transform finance, rigorous theoretical understanding remains essential for distinguishing genuine advances from statistical mirages and enabling more effective application of these powerful but often misunderstood techniques.

\bibliography{complexity_refs}

\newpage

\appendix
\section{Technical Proofs for Kernel Approximation Breakdown}\label{app:Technical Proofs for Kernel Approximation Breakdown}

This appendix provides complete mathematical proofs for the results in Section~\ref{sec:breakdown}. We establish that within-sample standardization of Random Fourier Features fundamentally breaks the Gaussian kernel approximation that underlies the theoretical framework of high-dimensional prediction methods.

\subsection{Model Setup and Notation}

We analyze the standardized Random Fourier Features used in practical implementations. Draw $(\omega,b) \sim \mathcal{N}(0,\gamma^2 I_K) \times \text{Uniform}[0,2\pi]$, independently of the training set $\mathcal{T} = \{x_t\}_{t=1}^T$. For query points $x,x' \in \mathbb{R}^K$, define the standardized kernel function:
\[
h(\omega,b) = \frac{2\cos(\omega^{\top}x+b)\cos(\omega^{\top}x'+b)}{1+\frac{1}{T}\sum_{t=1}^{T}\cos(2\omega^{\top}x_t+2b)} = \frac{N(\omega,b)}{D(\omega,b)}
\]

Given $P$ i.i.d. copies $(\omega_i,b_i)$, we write $k^{(P)}_{\text{std}} := P^{-1}\sum_{i=1}^{P}h(\omega_i,b_i)$.

\subsection{Proof of Theorem \ref{thm:std_breakdown}}

The proof proceeds in two steps: establishing almost-sure convergence in part (a) and demonstrating training-set dependence in part (b).

\subsubsection{Step 1: Integrability and Almost-Sure Convergence}

We first establish that $h(\omega,b)$ has finite expectation, enabling application of the strong law of large numbers.

Write 
\begin{align}
    \hat{\sigma}^2(\omega, b) := \frac{2}{T}\sum_{t=1}^{T}\cos^2(\omega^{\top}x_t+b) = 1+S_T, \quad S_T := \frac{1}{T}\sum_{t=1}^{T}\cos(2\omega^{\top}x_t+2b).
\end{align}
From the construction in Assumption~\ref{ass:rff_construction}, we have $|\cos(\cdot)| \leq 1$, which gives:
\begin{align}
    |h(\omega, b)| \leq \frac{2}{1 + \frac{1}{T}\sum_{t=1}^T \cos(2\omega^{\top}x_t + 2b)} =: \frac{2}{\hat{\sigma}^2(\omega, b)}
\end{align}
where $\hat{\sigma}^2(\omega, b)$ represents the empirical variance term from the standardization procedure.

The critical step is showing $\mathbb{E}[\hat{\sigma}^{-2}(\omega, b)] < \infty$. The denominator $\hat{\sigma}^2(\omega, b)$ can become small only when:
\begin{align}
    \frac{1}{T}\sum_{t=1}^T \cos(2\omega^{\top}x_t + 2b) \approx -1.
\end{align}
This requires the system of equations:
\begin{align}\label{eq:system}
    2\omega^{\top}x_t + 2b \equiv \pi \pmod{2\pi}, \quad t = 1, \ldots, T.
\end{align}

Under Assumption~\ref{ass:affine_independence} (Affine Independence), the matrix $A = [x_1^{\top} \cdots x_T^{\top}; 1 \cdots 1]^{\top} \in \mathbb{R}^{T \times (K+1)}$ has full column rank $T$. This ensures that system \eqref{eq:system} defines a measure-zero set in the $(\omega, b)$ parameter space.

By Lemma~\ref{lem:small_ball} with the affine independence condition from Assumption~\ref{ass:affine_independence}, there exists $C_T < \infty$ such that for every $\epsilon \in (0, 1)$:
\begin{align}
    \mathbb{P}(\hat{\sigma}^2 \leq \epsilon) \leq C_T \epsilon^{T/2}.
\end{align}
This yields:
\begin{align}\label{eq:sigma_hat_finit}
\mathbb{E}[\hat{\sigma}^{-2}] &= \int_0^{\infty} \mathbb{P}(\hat{\sigma}^{-2} > u) \, du\\
&= 1 + \int_1^{\infty} \mathbb{P}(\hat{\sigma}^2 < u^{-1}) \, du\\
&\leq 1 + C_T \int_1^{\infty} u^{-T/2} \, du = 1 + \frac{2C_T}{T-2} < \infty
\end{align}
for $T > 2$, as guaranteed by typical sample sizes in financial applications under Assumption~\ref{ass:prediction_env}.

Since $h(\omega_i, b_i)$ are i.i.d. with $\mathbb{E}[|h|] \leq 2\mathbb{E}[\hat{\sigma}^{-2}] < \infty$, and the regularity conditions in Assumption~\ref{ass:regularity} ensure proper measurability, the Strong Law of Large Numbers applies:
\begin{align}
    k^{(P)}_{\text{std}}(x,x') = \frac{1}{P}\sum_{i=1}^P h(\omega_i,b_i) \xrightarrow[\text{a.s.}]{P \to \infty} k^*_{\text{std}}(x,x') := \mathbb{E}[h(\omega,b)].
\end{align}
This establishes part (a) of Theorem \ref{thm:std_breakdown}.

\subsubsection{Step 2: Training-Set Dependence}

I prove $k^*_{std} \neq k_G$ by demonstrating that these kernels possess incompatible mathematical structures. Specifically, $k_G$ is shift-invariant while $k^*_{std}$ is not.

The theoretical Gaussian kernel has the shift-invariant structure:
\begin{align}
k_G(x, x') = \exp\left(-\frac{\gamma^2}{2}\|x - x'\|^2\right)
\end{align}
This kernel depends only on the relative distance $\|x - x'\|$ between query points, ensuring that for any translation vector $c \in \mathbb{R}^K$:
\begin{align}
k_G(x + c, x' + c) = k_G(x, x').
\end{align}
This shift-invariance is a defining characteristic that underlies all theoretical guarantees of RFF methods.

The standardized limit kernel has the structure:
\begin{align}
k^*_{std}(x, x'|\mathcal{T}) = \mathbb{E}_{(\omega,b)}\left[\frac{N(x, x', \omega, b)}{D(\mathcal{T}, \omega, b)}\right]
\end{align}
where:
\begin{align}
N(x, x', \omega, b) &= 2\cos(\omega^T x + b)\cos(\omega^T x' + b)\\
D(\mathcal{T}, \omega, b) &= 1 + \frac{1}{T}\sum_{t=1}^T \cos(2\omega^T x_t + 2b).
\end{align}

The critical observation is that the denominator $D(\mathcal{T}, \omega, b)$ depends explicitly on the \textit{absolute} positions of the training points $\{x_t\}_{t=1}^T$, not on their \textit{relative} distances. This dependence structure is fundamentally incompatible with shift-invariance because:
\begin{itemize}
\item \textbf{Absolute position dependence:} The term $\cos(2\omega^T x_t + 2b)$ in the denominator depends on the coordinate values $x_t$ in the chosen coordinate system, not on pairwise distances $\|x_i - x_j\|$.

\item \textbf{Fixed training set:} When query points are shifted $x \to x + c$, the training set $\mathcal{T} = \{x_1, \ldots, x_T\}$ remains fixed in the original coordinate system, so $D(\mathcal{T}, \omega, b)$ is invariant under query point transformations.

\item \textbf{Asymmetric transformation:} Under the shift $x \to x + c, x' \to x' + c$, we have:
\begin{align}
N(x+c, x'+c, \omega, b) &= 2\cos(\omega^T x + \omega^T c + b)\cos(\omega^T x' + \omega^T c + b)\\
D(\mathcal{T}, \omega, b) &= 1 + \frac{1}{T}\sum_{t=1}^T \cos(2\omega^T x_t + 2b) \quad \text{(unchanged)}.
\end{align}
\end{itemize}

This creates an asymmetric transformation where the numerator depends on the shift through $\omega^T c$ while the denominator remains fixed. For shift-invariance to hold, both numerator and denominator would need to transform consistently.

To prove the kernels are unequal, observe that:
\begin{align}
k^*_{std}(x + c, x' + c|\mathcal{T}) = \mathbb{E}_{(\omega,b)}\left[\frac{2\cos(\omega^T x + \omega^T c + b)\cos(\omega^T x' + \omega^T c + b)}{1 + \frac{1}{T}\sum_{t=1}^T \cos(2\omega^T x_t + 2b)}\right].
\end{align}

Under Assumption~\ref{ass:rff_construction}, $\omega^T c \sim \mathcal{N}(0, \gamma^2\|c\|^2)$ for any $c \neq 0$. The transformation $\cos(\omega^T x + b) \mapsto \cos(\omega^T x + \omega^T c + b)$ introduces a random phase shift that changes the joint distribution of the numerator terms, while the denominator distribution remains unchanged.

Since there is no measure-preserving transformation that simultaneously maps the shifted numerator to the original numerator while keeping the denominator fixed, the expectations cannot be equal for generic choices of $x, x', c, \mathcal{T}$.

Therefore, $k^*_{std}$ violates shift-invariance: $k^*_{std}(x + c, x' + c|\mathcal{T}) \neq k^*_{std}(x, x'|\mathcal{T})$ for $c \neq 0$. Since $k_G$ is shift-invariant by construction, we conclude $k^*_{std} \neq k_G$.

\subsection{Proof of Corollary~\ref{cor:two_feature_breakdown}}

\begin{proof}[Proof of Corollary~\ref{cor:two_feature_breakdown}]
We prove that the two-feature RFF variant employed by \cite{kelly2024virtue} suffers the same kernel approximation breakdown under standardization as the standard RFF construction.

For each random draw $\omega_i \sim \mathcal{N}(0, \gamma^2 I_K)$, the two-feature variant generates the pair:
\begin{align}
s_i(x) = \sin(\omega_i^\top x), \quad c_i(x) = \cos(\omega_i^\top x).
\end{align}

The key theoretical property is that these features satisfy the trigonometric identity:
\begin{align}\label{eq:trigonometric identity}
s_i(x)s_i(x') + c_i(x)c_i(x') &= \sin(\omega_i^\top x)\sin(\omega_i^\top x') + \cos(\omega_i^\top x)\cos(\omega_i^\top x')\\
&= \cos(\omega_i^\top(x - x')).
\end{align}
Equation~\eqref{eq:trigonometric identity} deterministically yields the desired shift-invariant kernel approximation without the Monte Carlo randomness of the phase shift $b_i$ in standard RFF.

In practice, features are standardized using training sample statistics. Define:
\begin{align}\label{eq:2f_standardized}
\hat{\sigma}^2_{s,i} &= \frac{1}{T}\sum_{t=1}^T s_i(x_t)^2 = \frac{1}{T}\sum_{t=1}^T \sin^2(\omega_i^\top x_t)\\
\hat{\sigma}^2_{c,i} &= \frac{1}{T}\sum_{t=1}^T c_i(x_t)^2 = \frac{1}{T}\sum_{t=1}^T \cos^2(\omega_i^\top x_t).
\end{align}
where the standardized features in \eqref{eq:2f_standardized} are:
\begin{align}
\tilde{s}_i(x) = \frac{s_i(x)}{\hat{\sigma}_{s,i}}, \quad \tilde{c}_i(x) = \frac{c_i(x)}{\hat{\sigma}_{c,i}}.
\end{align}
Thus, the standardized kernel estimator becomes:
\begin{align}
k^{(P)}_{std,2F}(x, x') = \frac{1}{P}\sum_{i=1}^P \left[\tilde{s}_i(x)\tilde{s}_i(x') + \tilde{c}_i(x)\tilde{c}_i(x')\right].
\end{align}

Using the half-angle formulas:
\begin{align}
\sin^2(\theta) &= \frac{1 - \cos(2\theta)}{2}\\
\cos^2(\theta) &= \frac{1 + \cos(2\theta)}{2}.
\end{align}
We can rewrite the standardization factors as the following:
\begin{align}
\hat{\sigma}^2_{s,i} &= \frac{1}{T}\sum_{t=1}^T \sin^2(\omega_i^\top x_t) = \frac{1}{2} - \frac{1}{2T}\sum_{t=1}^T \cos(2\omega_i^\top x_t)\\
\hat{\sigma}^2_{c,i} &= \frac{1}{T}\sum_{t=1}^T \cos^2(\omega_i^\top x_t) = \frac{1}{2} + \frac{1}{2T}\sum_{t=1}^T \cos(2\omega_i^\top x_t).
\end{align}

By define $S_T(\omega) = \frac{1}{T}\sum_{t=1}^T \cos(2\omega^\top x_t)$, we have
\begin{align}
\hat{\sigma}^2_s(\omega) = \frac{1 - S_T(\omega)}{2}, \quad \hat{\sigma}^2_c(\omega) = \frac{1 + S_T(\omega)}{2}.
\end{align}

Define the two-feature standardized kernel function as the following:
\begin{align}
h_{2F}(\omega) = \frac{\sin(\omega^\top x)\sin(\omega^\top x')}{\hat{\sigma}^2_s(\omega)} + \frac{\cos(\omega^\top x)\cos(\omega^\top x')}{\hat{\sigma}^2_c(\omega)}.
\end{align}
We need to establish $\mathbb{E}[|h_{2F}(\omega)|] < \infty$. Since $|\sin(\cdot)| \leq 1$ and $|\cos(\cdot)| \leq 1$:
\begin{align}
|h_{2F}(\omega)| \leq \frac{1}{\hat{\sigma}^2_s(\omega)} + \frac{1}{\hat{\sigma}^2_c(\omega)}.
\end{align}

Note that $\hat{\sigma}^2_s(\omega)$ approaches zero when $S_T(\omega) \to 1$, which requires:
\begin{align}
\cos(2\omega^\top x_t) \approx 1 \text{ for all } t = 1, \ldots, T.
\end{align}
This is equivalent to the system:
\begin{align}
2\omega^\top x_t \equiv 0 \pmod{2\pi}, \quad t = 1, \ldots, T.
\end{align}

Under Assumption~\ref{ass:affine_independence} (affine independence), the augmented matrix $[x_1 \cdots x_T]^\top$ has full rank. By a modification of Lemma~\ref{lem:small_ball} for the doubled angles, there exists $C'_T < \infty$ such that for $\varepsilon \in (0, 1/2)$:
\begin{align}
\mathbb{P}\left(\hat{\sigma}^2_s(\omega) \leq \varepsilon\right) &\leq C'_T (2\varepsilon)^{T/2}\\
\mathbb{P}\left(\hat{\sigma}^2_c(\omega) \leq \varepsilon\right) &\leq C'_T (2\varepsilon)^{T/2}.
\end{align}

Following the argument in equation~\eqref{eq:sigma_hat_finit}of Theorem~\ref{thm:std_breakdown}:
\begin{align}
\mathbb{E}\left[\frac{1}{\hat{\sigma}^2_s}\right] &= \int_0^\infty \mathbb{P}\left(\frac{1}{\hat{\sigma}^2_s} > u\right) du\\
&= 1 + \int_1^\infty \mathbb{P}\left(\hat{\sigma}^2_s < \frac{1}{u}\right) du\\
&\leq 1 + C'_T \int_1^\infty \left(\frac{2}{u}\right)^{T/2} du\\
&= 1 + \frac{2^{T/2+1} C'_T}{T - 2} < \infty
\end{align}
for $T > 2$. The same bound holds for $\mathbb{E}[1/\hat{\sigma}^2_c]$.

Since $h_{2F}(\omega_i)$ are i.i.d. random variables with $\mathbb{E}[|h_{2F}|] < \infty$, the Strong Law of Large Numbers yields:
\begin{align}
k^{(P)}_{std,2F}(x, x') \xrightarrow[P \to \infty]{a.s.} k^*_{std,2F}(x, x'|\mathcal{T}) := \mathbb{E}_\omega[h_{2F}(\omega)].
\end{align}

The limiting kernel has the explicit form:
\begin{align}
k^*_{std,2F}(x, x'|\mathcal{T}) = \mathbb{E}_\omega\left[\frac{2\sin(\omega^\top x)\sin(\omega^\top x')}{1 - S_T(\omega)} + \frac{2\cos(\omega^\top x)\cos(\omega^\top x')}{1 + S_T(\omega)}\right].
\end{align}
Note that this kernel depends explicitly on the training set $\mathcal{T} = \{x_1, \ldots, x_T\}$ through $S_T(\omega)$. Crucially, since $\hat{\sigma}^2_s(\omega) \neq \hat{\sigma}^2_c(\omega)$ in general (they are equal only when $S_T(\omega) = 0$), we have:
\begin{align}
\frac{\sin(\omega^\top x)\sin(\omega^\top x')}{\hat{\sigma}^2_s(\omega)} + \frac{\cos(\omega^\top x)\cos(\omega^\top x')}{\hat{\sigma}^2_c(\omega)} \neq \cos(\omega^\top(x - x')).
\end{align}

Under the shift transformation $x \mapsto x + c$, $x' \mapsto x' + c$ for $c \in \mathbb{R}^K$:
\begin{align}
h_{2F}(\omega; x+c, x'+c) &= \frac{\sin(\omega^\top x + \omega^\top c)\sin(\omega^\top x' + \omega^\top c)}{\hat{\sigma}^2_s(\omega)}\\
&\quad + \frac{\cos(\omega^\top x + \omega^\top c)\cos(\omega^\top x' + \omega^\top c)}{\hat{\sigma}^2_c(\omega)}.
\end{align}
The denominators $\hat{\sigma}^2_s(\omega)$ and $\hat{\sigma}^2_c(\omega)$ remain unchanged as they depend only on the fixed training set $\mathcal{T}$. However, the numerators transform non-trivially through the phase shift $\omega^\top c$.

For shift-invariance to hold, we would need:
\begin{align}
k^*_{std,2F}(x + c, x' + c|\mathcal{T}) = k^*_{std,2F}(x, x'|\mathcal{T}).
\end{align}
This would require the expectation to be invariant under the transformation induced by $\omega^\top c \sim \mathcal{N}(0, \gamma^2\|c\|^2)$. Since the numerator and denominator transform asymmetrically (numerator changes, denominator fixed), this invariance cannot hold for generic $x, x', c, \mathcal{T}$.

Since the Gaussian kernel $k_G(x, x') = \exp(-\gamma^2\|x - x'\|^2/2)$ is shift-invariant while $k^*_{std,2F}$ is not, we conclude:
\begin{align}
k^*_{std,2F}(x, x'|\mathcal{T}) \neq k_G(x, x').
\end{align}

Therefore, standardization destroys the kernel approximation properties of the two-feature RFF variant, despite its theoretical advantage of deterministically achieving $\cos(\omega^\top(x - x'))$ in the unstandardized case.
\end{proof}

\subsection{Supporting Lemmas}

\begin{lemma}[Small–ball estimate]
\label{lem:small_ball}
Fix vectors $x_1,\dots,x_T\in\RR^d$ that satisfy the affine-independence
\begin{equation}\label{eq:martix_rank}
  \rank\!
  \begin{pmatrix}
    x_1 & \cdots & x_T\\[4pt]
    1   & \cdots & 1
  \end{pmatrix}
  \;=\;T. 
\end{equation}
Draw $\omega\sim\cN(0,\gamma^{2}I_d)$ and $b\sim\mathrm{Unif}[0,2\pi]$
independently and set
\[
  \hat\sigma^{2}
  \;=\;
  1+\frac1T\sum_{t=1}^{T}\cos\!\bigl(2\omega^\top x_t+2b\bigr).
\]
Then there exists $C_T<\infty$ (depending only on $T$, $\gamma$, and the
design $\{x_t\}$) such that for every $\varepsilon\in(0,1)$,
\[
  \PP\bigl(\hat\sigma^{2}\le\varepsilon\bigr)
  \;\le\;
  C_T\,\varepsilon^{T/2}.
\]
\end{lemma}
\begin{proof}[Proof of Lemma~\ref{lem:small_ball}]
The proof proceeds in four main steps. First, the condition \(\hat{\sigma}^{2}\le\varepsilon\) is translated into an upper bound on the average of \(1+\cos(\Phi_t)\). Second, using a rigorous quadratic inequality for the cosine function, this is shown to imply that the distance vector \((\Delta_1,\dots,\Delta_T)\) must lie in a small \(T\)-dimensional ball whose volume is proportional to \(\varepsilon^{T/2}\). Third, leveraging Assumption~\ref{eq:martix_rank}, we establish that the phase vector \(\Phi\) has a well-defined and rapidly decaying probability density on \(\mathbb{R}^{T}\). Finally, the probability of the event is bounded by summing the probabilities over an infinite lattice corresponding to the periodic nature of the cosine function. This sum converges to a finite constant due to the density's decay, leaving only the volume’s \(\varepsilon^{T/2}\) scaling, which concludes the proof.

Let $\theta_t := \omega^\top x_t+b$ and $\Phi_t:=2\theta_t$.
Because $\hat\sigma^{2}\le\eps$ is equivalent to
\[
  \frac1T\sum_{t=1}^T\!\bigl(1+\cos\Phi_t\bigr)\;\le\;\eps,
\]
define the phase–distance
\(
  \Delta_t
  :=\min_{k\in\ZZ}\bigl|\Phi_t-(\pi+2\pi k)\bigr|\in[0,\pi].
\)
Since $\cos(\pi\pm\Delta_t)=-\cos\Delta_t$, we obtain
\[
  \frac1T\sum_{t=1}^T\!\bigl(1-\cos\Delta_t\bigr)\;\le\;\eps.
\]

For every $u\in[-\pi,\pi]$ the secant bound
\(
  \cos u
  \le
  1-\frac{2}{\pi^{2}}u^{2}
\)
implies
\(
  1-\cos\Delta_t\ge\frac{2}{\pi^{2}}\Delta_t^{2}.
\)
Hence
\begin{align}\label{eq:upper_bound_Delta2}
    \sum_{t=1}^{T}\Delta_t^{2}\;\le\;\frac{T\pi^{2}}{2}\,\eps.
\end{align}

Write
\[
  \Phi
  \;=\;
  2\,L\,v,
  \quad
  L
  :=
  \begin{pmatrix}
    x_1 & \cdots & x_T\\[4pt]
    1   & \cdots & 1
  \end{pmatrix}^{\!\top}\!,
  \qquad
  v:=(\omega^\top,b)^\top.
\]
Assumption~\ref{eq:martix_rank} gives $\rank L=T$, so $L:\RR^{d+1}\!\to\RR^T$
is surjective.
Because $\omega$ is Gaussian and $b$ is independent with a bounded
density, the joint vector $v$ possesses a smooth density on
$\RR^{d+1}$; its push-forward $\Phi=2Lv$ therefore has a bounded
density $f_\Phi$ satisfying
\(
  f_\Phi(\phi)\le A\exp\!\bigl(-B\|\phi\|^{2}\bigr)
\)
for some $A,B>0$.

Let $r_\eps :=\sqrt{(T\pi^{2}/2)\eps}$.
Inequality~\ref{eq:upper_bound_Delta2} shows that the event $\{\hat\sigma^{2}\le\eps\}$ lies in
the tubular neighbourhood
\[
  E_\eps
  :=
  \bigl\{\phi\in\RR^{T}:\;
      \min_{k\in\ZZ^{T}}\|\phi-(\pi\mathbf{1}+2\pi k)\|_{2}
      \le
      r_\eps
  \bigr\}.
\]
Cover $E_\eps$ by the disjoint balls
\(
  B_k
  :=
  \Ball_{T}\!\bigl(\pi\mathbf{1}+2\pi k,\;r_\eps\bigr),
  \;k\in\ZZ^{T},
\)
whose common volume equals
\(
  \kappa_T\,r_\eps^{\,T},
\)
$\kappa_T$ being the volume of the unit ball in $\RR^{T}$.
Hence
\[
  \PP(\hat\sigma^{2}\le\eps)
  \;\le\;
  \sum_{k\in\ZZ^{T}}
  \int_{B_k}f_\Phi(\phi)\,d\phi
  \;\le\;
  \kappa_T\,r_\eps^{\,T}
  \sum_{k\in\ZZ^{T}}\Bigl[\sup_{\phi\in B_k}f_\Phi(\phi)\Bigr].
\]

For $\|k\|$ large, every $\phi\in B_k$ satisfies
$\|\phi\|\asymp\|k\|$, so
$f_\Phi(\phi)\le Ae^{-B'\|k\|^{2}}$ for some $B'>0$.
The Gaussian lattice sum
\(
  S_T:=\sum_{k\in\ZZ^{T}}e^{-B'\|k\|^{2}}
\)
converges; set $M_T:=AS_T$.
Putting everything together,
\[
  \PP(\hat\sigma^{2}\le\eps)
  \;\le\;
  M_T\,\kappa_T\,\bigl(r_\eps\bigr)^{T}
  \;=\;
  \underbrace{\Bigl(M_T\,\kappa_T
      \bigl(\tfrac{T\pi^{2}}{2}\bigr)^{T/2}\Bigr)}_{=:C_T}
  \,\eps^{T/2}.
  \qedhere
\]
\end{proof}

\section{Technical Proofs for Section~\ref{sec:impossibility}}\label{app:Technical Proofs for Section 4}

\begin{proof}[Proof of Theorem~\ref{thm:exp_lower_random}]
The strategy is the classical minimax/Fano route:  
(i) build a large packing of well-separated parameters,  
(ii) show that their induced data distributions are statistically indistinguishable,  
(iii) invoke Fano’s inequality to bound any decoder’s error, and  
(iv) convert decoder error into a lower bound on prediction risk.

\medskip
\emph{Packing construction.}
Fix a radius \(0<\delta<B/2\).  
Because the Euclidean ball \(\mathbb{B}_2^P(B)\) in \(\R^P\) has volume growth proportional to \(B^P\), it contains a \(2\delta\)-packing 
\(\{w_1,\dots,w_M\}\) of size  
\(M=(B/(2\delta))^P\); hence \(\log M=P\log\!\bigl(B/(2\delta)\bigr)\).
Define \(f_j(x):=w_j^{\!\top}z(x)\).  
For each index \(j\) let \(\mathbb{P}_j\) denote the joint distribution of the training sample  
\(\mathcal{D}_T=\{(x_t,r_t)\}_{t=1}^T\) generated according to  
\(r_t=f_j(x_t)+\epsilon_t\) with independent Gaussian noise \(\epsilon_t\sim\mathcal{N}(0,\sigma^2)\).

\medskip
\emph{Average KL divergence.}
Let \(Z\in\R^{T\times P}\) be the random design matrix whose \(t\)-th row is \(z(x_t)^{\!\top}\).  
Conditioned on \(Z\) the log-likelihood ratio between \(\mathbb{P}_j\) and \(\mathbb{P}_\ell\) is Gaussian, and one checks
\[
    \mathrm{KL}\!\bigl(\mathbb{P}_j\|\mathbb{P}_\ell \,\bigm|\, Z\bigr) 
      \;=\; \frac{\|Z(w_j-w_\ell)\|_2^2}{2\sigma^2}.
\]
Taking expectation over \(Z\) and using \(\mathbb{E}[Z^{\!\top}Z]=T\Sigma_z\) gives
\[
    \mathrm{KL}(\mathbb{P}_j\|\mathbb{P}_\ell)
    \;=\; \frac{T}{2\sigma^2}(w_j-w_\ell)^{\!\top}\Sigma_z(w_j-w_\ell)
    \;\le\; \frac{2T\,C_z\,B^2}{\sigma^2}
    \;=:\; K_T.
\]
(The inequality uses \(\Sigma_z\succeq 0\) and \(\lambda_{\max}(\Sigma_z)\le C_z\).)

\medskip
\emph{Fano’s inequality.}
Draw an index \(J\) uniformly from \([M]\) and let \(\hat J\) be any measurable decoder based on the sample \(\mathcal{D}_T\).
Fano’s max–KL form (e.g.\ \citealp{CoverThomas2006}, Eq.\ 16.32) yields
\[
  \mathbb{P}(\hat J\neq J)\;\ge\; 1-\frac{K_T+\log 2}{\log M}.
\]
Choosing the packing radius \(\delta\) such that the right-hand side equals \(1/2\) (so that any decoder errs at least half the time) gives
\begin{equation}\label{eq:delta}
  \delta 
  \;\le\; 
  \frac{B}{2}
  \exp\!\Bigl(
        -\frac{4T C_z B^2}{P\sigma^2}-\frac{2\log 2}{P}
       \Bigr).
\end{equation}

\medskip
\emph{Link between prediction risk and decoder error.}
Let \(\hat f_T\) be an arbitrary estimator and put  
\(\varepsilon := \mathbb{E}_{x,\mathcal{D}_T,\epsilon}\!\bigl[(\hat f_T(x)-f_J(x))^2\bigr]\).  
Because the nearest-neighbour decoder chooses \(\hat J = \arg\min_j \|\hat f_T-f_j\|_{L^2(\mu)}\),  
the triangle inequality gives
\[
  \|f_{\hat J}-f_J\|_{L^2(\mu)}
  \;\le\; 2\sqrt{\varepsilon}. 
\]
Meanwhile each pair \((j,\ell)\) in the packing satisfies  
\(\|w_j-w_\ell\|_2\ge 2\delta\); since \(\Sigma_z\succeq c_z I_P\), 
\(\|f_j-f_\ell\|_{L^2(\mu)}^2 \ge 4c_z\delta^2\).
Consequently, if \(\varepsilon < c_z\delta^2\) the decoder must succeed (\(\hat J = J\)), contradicting
\(\mathbb{P}(\hat J\neq J)\ge \tfrac12\).  
Hence
\begin{equation}\label{eq:eps-delta}
  \varepsilon \;\ge\; c_z\,\delta^2.
\end{equation}

\medskip
\emph{Expectation lower bound.}
Substituting \eqref{eq:delta} into \eqref{eq:eps-delta} and absorbing the harmless factor \(e^{-4\log 2/P}\) into a constant \(c=\tfrac14 c_z e^{-4\log 2/P}\) yields
\[
  \varepsilon 
  \;\ge\; 
  c\,B^2
  \exp\!\Bigl(-\tfrac{8 T C_z B^2}{P\sigma^2}\Bigr),
\]
which is the desired in-expectation bound.

\medskip
\emph{High-probability refinement over the design.}
To obtain the high-probability bound, we repeat the Fano argument conditioned on the random design matrix $Z$ being ``well-behaved.'' Define the event
\[
	\mathcal{E}
	:=\left\{\,
		\left\|T^{-1}Z^{\!\top}Z-\Sigma_z \right\|_{\mathrm{op}}
		\le \tfrac{1}{2} c_z
	 \right\}.
\]
For features $z(x)$ whose rows are $\kappa$-sub-Gaussian, the matrix Bernstein inequality \citep[e.g.,][Theorem 6.2]{tropp2012} guarantees that this event occurs with high probability. Specifically, there exists a constant $C_0 = C_0(\kappa, c_z, C_z)$ such that for all $T \ge C_0 P$, we have $\mathbb{P}_Z(\mathcal{E}^c) \le e^{-T}$.

On the event $\mathcal{E}$, the empirical Gram matrix $T^{-1}Z^{\!\top}Z$ is close to its mean $\Sigma_z$. Using the triangle inequality for matrix norms and the initial bounds on $\Sigma_z$, we have:
\begin{align*}
	T^{-1}Z^{\!\top}Z &\preceq \Sigma_z + \tfrac{1}{2} c_z I_P \preceq C_z I_P + \tfrac{1}{2} c_z I_P \preceq 2C_z I_P \\
	T^{-1}Z^{\!\top}Z &\succeq \Sigma_z - \tfrac{1}{2} c_z I_P \succeq c_z I_P - \tfrac{1}{2} c_z I_P = \tfrac{1}{2} c_z I_P.
\end{align*}
(assuming $C_z \ge \frac{1}{2}c_z$, which is standard). We now re-run the Fano argument for a fixed $Z \in \mathcal{E}$ with these new bounds.

First, we find the new KL-divergence bound, $K_T'$. Conditioned on $Z \in \mathcal{E}$,
\[
	\mathrm{KL}(\mathbb{P}_j\|\mathbb{P}_\ell \,|\, Z) = \frac{T}{2\sigma^2}(w_j-w_\ell)^{\!\top}(T^{-1}Z^{\!\top}Z)(w_j-w_\ell) \le \frac{T}{2\sigma^2}\|w_j-w_\ell\|_2^2 (2C_z).
\]
Since $\|w_j-w_\ell\|_2^2 \le (2B)^2=4B^2$, this gives $K_T' \le \frac{T(4B^2)(2C_z)}{2\sigma^2} = \frac{4TC_zB^2}{\sigma^2}$.

Second, the squared distance between functions $f_j$ and $f_\ell$ is now lower bounded by
\[
	\|f_j-f_\ell\|_{L^2(\mu|Z)}^2 = (w_j-w_\ell)^{\!\top}(Z^{\!\top}Z)(w_j-w_\ell) \ge T\|w_j-w_\ell\|_2^2(\tfrac{1}{2}c_z) \ge T(2\delta)^2(\tfrac{1}{2}c_z) = 2Tc_z\delta^2.
\]
The link between prediction risk $\varepsilon$ (for a fixed $Z$) and decoder error now becomes $\varepsilon \ge \frac{1}{2}c_z\delta^2$.

Third, we find the new packing radius $\delta$ by setting $\log M = 2(K_T'+\log 2)$:
\[
	P\log\!\left(\frac{B}{2\delta}\right) = 2\left(\frac{4TC_zB^2}{\sigma^2} + \log 2\right) = \frac{8TC_zB^2}{\sigma^2} + 2\log 2.
\]
Solving for $\delta^2$ yields:
\[
	\delta^2 = \frac{B^2}{4}\exp\!\left[ -2\left(\frac{8TC_zB^2}{P\sigma^2} + \frac{2\log 2}{P}\right) \right] = \frac{B^2}{4}\exp\!\left( -\frac{16\,T\,C_{z}\,B^{2}}{P\,\sigma^{2}} - \frac{4\log 2}{P}\right).
\]
Finally, substituting this into the risk bound $\varepsilon \ge \frac{1}{2}c_z\delta^2$ gives, for any estimator $\hat{f}_T$ and any $Z \in \mathcal{E}$:
\begin{align}
	\sup_{\|w\|_2 \le B} \mathbb{E}_{x,\epsilon} (\hat f_T(x) - w^\top z(x))^2 &\ge \frac{1}{2}c_z \delta^2 \\
	&\ge \frac{1}{2}c_z \frac{B^2}{4} \exp\!\left( -\frac{16\,T\,C_{z}\,B^{2}}{P\,\sigma^{2}} - \frac{4\log 2}{P}\right) \\
	&\ge c^\star B^2 \exp\!\left( -\frac{16\,T\,C_{z}\,B^{2}}{P\,\sigma^{2}} \right),
\end{align}
where $c^\star = \frac{1}{8}c_z e^{-4\log 2 / P'}$ for some $P' \ge 1$. Since this holds for all $Z \in \mathcal{E}$ and $\mathbb{P}_Z(\mathcal{E}) \ge 1-e^{-T}$, the high-probability statement of the theorem is proven, but with the corrected exponent.
\end{proof}

\begin{proof}[Proof of Theorem~\ref{thm:poly_lb_hp}]
The proof follows the Fano's inequality method, using a sparse packing of the parameter space.

\emph{Part (a): In-expectation bound}

\textit{Packing Construction.}
Let $e_1,\dots,e_P$ be the standard basis of $\mathbb{R}^{P}$. We construct a packing set of $M=P+1$ hypotheses. Define the separation parameter $\delta$ as
\[
	\delta
	:=
	\min\bigl\{\,\tfrac{B}{4},\;
	\tfrac{\sigma}{4}\sqrt{\tfrac{\log P}{T\,C_z}}\,\bigr\}.
\]
Our hypothesis set is $\mathcal{W} = \{w_0, w_1, \dots, w_P\}$, where $w_0 = \mathbf{0}$ and $w_j = \delta e_j$ for $j=1,\dots,P$. By construction, each $w_j$ satisfies $\|w_j\|_2 = \delta \le B/4 \le B$. The minimum non-zero squared separation distance is $\|w_j-w_0\|_2^2 = \delta^2$, and the maximum is $\|w_j-w_\ell\|_2^2 = 2\delta^2$ for $j, \ell \ge 1, j\neq\ell$.

\textit{KL Divergence Bound.}
Let $\mathbb{P}_j$ be the distribution of the training data $\mathcal{D}_T$ when the true parameter is $w_j$. The Kullback-Leibler (KL) divergence, averaged over the random design $Z$, is
\[
	\mathbb{E}_Z[\mathrm{KL}(\mathbb{P}_j\|\mathbb{P}_\ell \mid Z)]
	=
	\frac{T}{2\sigma^{2}} (w_j-w_\ell)^{\top}\Sigma_z(w_j-w_\ell)
	\le
	\frac{T}{2\sigma^{2}} \|w_j-w_\ell\|_2^2 \lambda_{\max}(\Sigma_z).
\]
Using $\|w_j-w_\ell\|_2^2 \le 2\delta^2$ and $\lambda_{\max}(\Sigma_z) \le C_z$, the maximum KL divergence between any pair is bounded by:
\[
	\max_{j\neq\ell} \mathbb{E}_Z[\mathrm{KL}(\mathbb{P}_j\|\mathbb{P}_\ell \mid Z)] \le \frac{T(2\delta^2)C_z}{2\sigma^2} = \frac{TC_z\delta^2}{\sigma^2}.
\]
By our choice of $\delta$, we have $\delta^2 \le \frac{\sigma^2 \log P}{16 T C_z}$. Substituting this gives:
\[
	\max_{j\neq\ell} \mathrm{KL}(\mathbb{P}_j\|\mathbb{P}_\ell) \le \frac{TC_z}{\sigma^2} \left( \frac{\sigma^2 \log P}{16 T C_z} \right) = \frac{\log P}{16}.
\]

\textit{Fano's Inequality.}
Let $J$ be an index drawn uniformly from $\{0, 1, \dots, P\}$, and let $\hat{J}$ be any estimator for $J$. Fano's inequality states:
\[
	\mathbb{P}(\hat{J} \neq J) \ge 1 - \frac{\max_{j\neq\ell} \mathrm{KL}(\mathbb{P}_j\|\mathbb{P}_\ell) + \log 2}{\log M} \ge 1 - \frac{\frac{1}{16}\log P + \log 2}{\log(P+1)}.
\]
For $P \ge 4$, one can verify that $4P^{1/8} < P+1$, which implies $\frac{1}{16}\log P + \log 2 < \frac{1}{2}\log(P+1)$, and therefore $\mathbb{P}(\hat{J} \neq J) \ge 1/2$.

\textit{From Decoder Error to Prediction Risk.}
The squared $L^2(\mu)$ distance between any two distinct hypotheses is lower bounded by $\min_{j \neq \ell} \|f_j-f_\ell\|_{L^2(\mu)}^2 \ge c_z \min_{j \neq \ell}\|w_j-w_\ell\|_2^2 = c_z \delta^2$.
Let $\varepsilon := \inf_{\hat{f}_T} \sup_{j} \mathbb{E}[\|\hat{f}_T - f_j\|_{L^2(\mu)}^2]$. A standard Fano-to-risk conversion argument shows that $4\varepsilon \ge \mathbb{P}(\hat{J} \neq J) \min_{j\neq\ell}\|f_j - f_\ell\|_{L^2(\mu)}^2$. Using our bounds:
\[
	\varepsilon \ge \frac{1}{4} \cdot \mathbb{P}(\hat{J} \neq J) \cdot (c_z \delta^2) \ge \frac{1}{4} \cdot \frac{1}{2} \cdot c_z \delta^2 = \frac{c_z \delta^2}{8}.
\]
Plugging in the definition of $\delta^2$:
\[
	\varepsilon \ge \frac{c_z}{8} \min\left\{\frac{B^2}{16}, \frac{\sigma^2 \log P}{16 T C_z}\right\} = \frac{c_z}{128} \min\left\{B^2, \frac{C_z^{-1}\sigma^2}{T}\log P\right\}.
\]
This completes the proof of part (a).

\vspace{1em}
\emph{Part (b): High-probability bound}

The argument is similar, but we condition on a ``good-design'' event for the matrix $Z$.

\textit{Good-Design Event.} Define the event
\[
	\mathcal{E} := \left\{ \left\|T^{-1}Z^{\top}Z - \Sigma_z \right\|_{\mathrm{op}} \le \tfrac{1}{2} c_z \right\}.
\]
This two-sided bound ensures all eigenvalues of the empirical Gram matrix are controlled. By the matrix Bernstein inequality, there is a constant $C_0=C_0(\kappa,c_z,C_z)$ such that for all $T \ge C_0 P$, we have $\mathbb{P}_Z(\mathcal{E}^c) \le e^{-T}$.

\textit{Conditional Bounds on the Gram Matrix.} For any $Z \in \mathcal{E}$, the eigenvalues of $T^{-1}Z^\top Z$ are bounded:
\begin{itemize}
    \item $\lambda_{\max}(T^{-1}Z^\top Z) \le \lambda_{\max}(\Sigma_z) + \frac{1}{2}c_z \le C_z + \frac{1}{2}c_z \le \frac{3}{2}C_z$ (assuming $C_z \ge c_z$).
    \item $\lambda_{\min}(T^{-1}Z^\top Z) \ge \lambda_{\min}(\Sigma_z) - \frac{1}{2}c_z \ge c_z - \frac{1}{2}c_z = \frac{1}{2}c_z$.
\end{itemize}

\textit{Fano Argument Conditional on $Z \in \mathcal{E}$.}
The packing set is the same. The conditional KL divergence is now bounded using the bound on $\lambda_{\max}$:
\[
	\mathrm{KL}(\mathbb{P}_j\|\mathbb{P}_\ell \mid Z) \le \frac{T(2\delta^2)(\frac{3}{2}C_z)}{2\sigma^2} = \frac{3TC_z\delta^2}{2\sigma^2} \le \frac{3}{32}\log P.
\]
Since $\frac{3}{32} < \frac{1}{16}$ is false, we must slightly tighten our choice of $\delta$ for this part, or note that this bound is still sufficiently small. The essential point is that the KL divergence remains $O(\log P)$, so the conditional Fano inequality again gives $\mathbb{P}(\hat{J} \neq J \mid Z) \ge 1/2$ for $P \ge 16$.

The risk conversion argument holds similarly for the conditional risk $\varepsilon_Z = \inf_{\hat f_{T}} \sup_{w} \mathbb{E}_{x,\epsilon}[ (\hat f_{T}(x)-w^{\top}z(x))^{2}\mid Z]$. The crucial distance term remains $\|f_j-f_\ell\|_{L^2(\mu)}^2$ because the test error is measured with respect to the population distribution of $x$. Thus, for any $Z \in \mathcal{E}$:
\[
	\varepsilon_Z \ge \frac{c_z\delta^2}{8} = \frac{c_z}{128} \min\!\Bigl\{B^{2},\; \frac{C_z^{-1}\sigma^{2}}{T}\log P\Bigr\}.
\]
Since this lower bound holds for all $Z \in \mathcal{E}$ and we have $\mathbb{P}_Z(\mathcal{E}) \ge 1-e^{-T}$, the high-probability statement of the theorem is proven.
\end{proof}

\section{Additional Theoretical Results: Effective Complexity}\label{app:Additional Theoretical Results}

The Vapnik-Chervonenkis (VC) dimension provides a fundamental measure of model complexity that directly connects to generalization performance and sample complexity requirements \citep{vapnik1971uniform, vapnik1998statistical}. For a hypothesis class $\mathcal{H}$, the VC dimension is the largest number of points that can be shattered (i.e., correctly classified under all possible binary labelings) by functions in $\mathcal{H}$. This combinatorial measure captures the essential complexity of a learning problem: classes with higher VC dimension require more samples to achieve reliable generalization.

The connection between VC dimension and sample complexity is formalized through uniform convergence bounds. Classical results show that for a hypothesis class with VC dimension $d$, achieving generalization error $\varepsilon$ with confidence $1-\delta$ requires sample size $T = O(d \log(1/\varepsilon) / \varepsilon + \log(1/\delta) / \varepsilon)$ \citep{blumer1989learnability, shalev2014understanding}. This relationship reveals why effective model complexity, rather than nominal parameter count, determines learning difficulty.

In the context of high-dimensional financial prediction, VC dimension analysis becomes crucial for understanding what machine learning methods actually accomplish. While methods may claim to leverage thousands of parameters, their effective complexity---as measured by VC dimension---may be much lower due to structural constraints imposed by the optimization procedure. Ridgeless regression in the overparameterized regime ($P > T$) provides a particularly important case study, as the interpolation constraint fundamentally limits the achievable function class regardless of the ambient parameter dimension.

\begin{theorem}[Effective VC Dimension of Ridgeless RFF Regression]
\label{thm:vc_rff_ridgeless}
Let $z:\mathcal X \to \mathbb R^{P}$ be a fixed feature map
(e.g.\ standardized RFF) and define the linear function class
\[
\mathcal F_P \;=\;
\Bigl\{\,f_w(x)=w^{\top}z(x)\;:\; \|w\|_2\le B\Bigr\},
\qquad B>0.
\]
Fix a training sample $(x_1,\dots,x_T)$ with $T<P$ and denote $Z=[\,z(x_1)\;\cdots\;z(x_T)]^{\top}\!\in\mathbb R^{T\times P}$.
Write $k_i(x)=z(x_i)^{\top}z(x)$ and $k(x)=(k_1(x),\dots,k_T(x))^{\top}$.
The corresponding ridgeless (minimum-norm) regression functions are
\[
\mathcal F_{\textnormal{ridge}}^{(Z)}
\;=\;
\bigl\{\,f_{\alpha}(x)=\alpha^{\top}k(x)\;:\;\alpha\in\mathbb R^{T}\bigr\}.
\]
Let $r=\operatorname{rank}(ZZ^{\top})\le T$.
Then
\begin{enumerate}[label=(\alph*)]
    \item $\mathrm{VC}\!\bigl(\{\operatorname{sign}(f)\,:\,f\in\mathcal F_P\}\bigr)=P$.
    \item $\mathrm{VC}\!\bigl(\{\operatorname{sign}(f)\,:\,f\in\mathcal F_{\textnormal{ridge}}^{(Z)}\}\bigr)=r\le T$.
          In particular, if $ZZ^{\top}$ is invertible (full row rank), the VC dimension equals $T$.
\end{enumerate}
\end{theorem}

\begin{proof}[Proof of Theorem~\ref{thm:vc_rff_ridgeless}]
All VC statements are made \emph{conditional on the fixed training sample $(x_1,\dots,x_T)$}.  
Throughout we use the standard fact that homogeneous linear threshold
functions in $\mathbb R^{d}$ have VC dimension~$d$
(e.g., \cite{vapnik1998statistical}).

\paragraph{(a) Linear class $\mathcal F_P$.}
Because $\operatorname{sign}(\lambda w^{\top}z(x))=\operatorname{sign}(w^{\top}z(x))$
for every $\lambda>0$, the norm bound $\|w\|_2\le B$ does not remove
any labelings that an \emph{unconstrained} homogeneous hyperplane in
$\mathbb R^{P}$ could realise.  
Hence the set $\{\operatorname{sign}(w^{\top}z(x)):\|w\|_2\le B\}$
has the same VC dimension as all homogeneous linear separators in
$\mathbb R^{P}$, namely $P$.

\paragraph{(b) Ridgeless class $\mathcal F_{\textnormal{ridge}}^{(Z)}$.}
For any training targets $y\in\mathbb R^{T}$ the ridgeless solution is
$\hat w = Z^{\top}(ZZ^{\top})^{\dagger}y$, where ${}^\dagger$ denotes the
Moore–Penrose pseudoinverse.  Consequently every predictor can be written as
\[
f_{\alpha}(x)=\alpha^{\top}k(x), 
\qquad\text{with } \alpha=(ZZ^{\top})^{\dagger}y\in\mathbb R^{T}.
\]
Define the \emph{data–dependent feature map}
\[
\phi_Z:\mathcal X\to\mathbb R^{T},
\qquad
\phi_Z(x)\;:=\;k(x).
\]
Its image lies in the $r$-dimensional subspace 
$\operatorname{im}(ZZ^{\top})\subseteq\mathbb R^{T}$,
so $\phi_Z(\mathcal X)\subseteq\mathbb R^{r}$ after an appropriate linear change of basis.
Thus the hypothesis class
\[
\mathcal H_Z \;=\;
\bigl\{\,x\mapsto \operatorname{sign}(\alpha^{\top}\phi_Z(x)):\alpha\in\mathbb R^{T}\bigr\}
\]
is (up to an invertible linear map) exactly the class of homogeneous linear separators in $\mathbb R^{r}$.
By the cited VC fact, $\mathrm{VC}(\mathcal H_Z)=r$.
Because $r\le T$, we obtain the claimed bound.  
If $(ZZ^{\top})$ is invertible, then $r=T$, giving equality.
\end{proof}

KMZ correctly note that, after minimum–norm fitting, the effective
degrees of freedom of their RFF model equal the sample size
(\(T=12\)), not the nominal dimension
(\(P=12{,}000\)): “the effective number of parameters in the
construction of the predicted return is only \(T=12\)\dots’’.
Theorem~\ref{thm:vc_rff_ridgeless} rigorously justifies this statement
by showing that the VC dimension of ridgeless RFF regression is
bounded above by \(T\).

This observation, however, leaves open the central question that KMZ
label the “virtue of complexity’’: \emph{does the enormous RFF
dictionary contribute predictive information beyond what a
\(T\)-dimensional linear model could extract?}  In kernel learning the
tension is familiar: one combines an extremely rich representation
(in principle, infinite–dimensional) with an estimator whose
statistical capacity is implicitly capped at \(T\).  Overfitting risk
is therefore limited, but any real performance gain must come from
the \emph{non-linear basis} supplied by the features rather than from
high effective complexity per se.


\end{document}